\documentclass[12pt]{article}

\newcommand{\be}{\begin{equation}}
\newcommand{\ee}{\end{equation}}
\usepackage{epsfig,amsmath,amssymb,graphics,graphicx} 

\usepackage{amscd}
\usepackage{pb-diagram}
\usepackage{hyperref}
\usepackage{amsfonts}
\usepackage{epsfig}

\begin{document}

\begin{center}

{\it Journal of Statistical Mechanics.
Vol.2014. No.9. (2014) P09036.}
\vskip 3mm

{\bf \large Large Lattice Fractional Fokker-Planck Equation} \\
\vskip 3mm

\vskip 7mm
{\bf \large Vasily E. Tarasov} \\
\vskip 3mm

{\it Skobeltsyn Institute of Nuclear Physics,\\ 
Lomonosov Moscow State University, \\
Moscow 119991, Russia} \\
{E-mail: tarasov@theory.sinp.msu.ru} \\

\begin{abstract}
Equation of long-range particle drift and diffusion 
on three-dimensional physical lattice is suggested.
This equation can be considered as a lattice analog
of space-fractional Fokker-Planck equation for continuum.
The lattice approach gives a possible microstructural basis 
for anomalous diffusion in media 
that are characterized by the non-locality of power-law type. 
In continuum limit the suggested three-dimensional lattice 
Fokker-Planck equations give fractional Fokker-Planck equations 
for continuous media with power-law non-locality
that is described by derivatives of non-integer orders.
The consistent derivation of 
the fractional Fokker-Planck equation 
is proposed as a new basis to describe 
space-fractional diffusion processes.
\end{abstract}

\end{center}



\section{Introduction}

Fokker-Planck equations are usually used to describe 
the Brownian motion of particles \cite{Risken}.
These equations describe the change of 
probability of a random function in space and time
in diffusion processes.
The Fokker-Planck equation is usually the 
second-order partial differential equation of parabolic type.
In many studies of diffusion processes in complex media,
the usual second-order Fokker-Planck equation 
may not be adequate. 
In particular, the probability density may have 
a thicker tail than the Gaussian probability density, and correspondent 
correlation functions may decay to zero much slower 
than the functions for usual diffusion processes 
resulting in long-range dependence. 
This phenomenon is known as anomalous diffusion 
\cite{BG1990,SZK1993,MK2000}.
Anomalous diffusion processes can be characterized 
by a power-law mean squared displacement of 
the form \cite{BG1990,SZK1993,MK2000}
\be \label{r2}
\langle x^2(t) \rangle = 
\frac{2 \, K (\alpha) t^{\alpha}}{\Gamma (\alpha +1)} ,
\ee
where $\Gamma (z)$ is the Gamma function, 
$\alpha$ is the anomalous diffusion exponent, 
and $K(\alpha)$ is the anomalous diffusion constant.
In equation (\ref{r2}), we use the second moment 
that is defined in terms of the ensemble average.
Depending on the value of $\alpha$, we 
usually distinguish sub-diffusion 
for $0 < \alpha < 1$ or super-diffusion for $\alpha > 1$. 
There are two limit cases such as 
the normal diffusion ($\alpha = 1$) 
and the ballistic motion ($\alpha = 2$).
One of possible approach to describe 
the anomalous diffusion is based 
on the continuous time random walk models \cite{Hughes} 
in which the particles are considered 
as random walkers with step lengths ${\bf r}$ 
and waiting times $t$.
An important role play the anomalous diffusion processes
with the Poissonian waiting time and 
the L\'evy distribution for the jump length.
The L\'evy flights \cite{BG1990} 
are random walks in which the step-lengths (long jumps)
have a probability distribution that is heavy-tailed.
The L\'evy motion can be described by generalized diffusion 
equation with space derivatives of non-integer orders $\mu$, 
\cite{MK2000}.
The fractional moment of order $\delta$ 
for L\'evy flights has the form
$\langle |x(t)|^{\delta} \rangle \sim t^{\delta / \mu}$,
where $0< \delta <\mu \le 2$.

Derivatives of non-integer orders 
\cite{SKM,KST,Ort,Uch,Zhou,Mainardi1997,FC1,FC2} 
play an important role in describing 
particle transport in anomalous diffusion
\cite{MK2000,MLP2001,Zaslavsky2002,MK2004,KLM,MS2012,US} 
and have a wide application in various areas of physics
(see for example \cite{CM,Hilfer,SATM,LA,Mainardi,TarasovSpringer,IJMPB2013,APSS2014a,APSS2014b}).
Various approaches lead to different types of 
space-time fractional Fokker-Planck equations. 
Usually the space-fractional Fokker-Planck equations
are obtained from the second-order differential equations
by replacing the first-order and second-order space 
derivatives by fractional-order derivatives \cite{KST}. 
Fractional Fokker-Planck equations with coordinate 
derivatives of non-integer order 
have been suggested in \cite{Zaslavsky1994}. 
The solutions and properties of these
equations are described in \cite{SZ1997,Zaslavsky2002}.
The Fokker-Planck equation with fractional coordinate derivatives
was also considered in \cite{Milovanov2001,MLP2001,MN2002,Chaos2005,Physica2008,TSMD2012}.
It should be noted that the fractional Fokker-Planck equations 
can be derived from the probabilistic continuous
time random walk \cite{MBK-1,MBK-2,BMK}.
In this paper we propose a consistent derivation of
the space-fractional Fokker-Planck equation 
based on lattice model with long-range drift and diffusion 
that is considered as a microstructural basis
to describe fractional diffusion processes in continua.

A discrete lattice version of the Fokker-Planck equation 
in analogy with the lattice-Boltzmann models
has been suggested in \cite{LFPE-1,LFPE-2,LFPE-3}.
These models are used 
to solve the equations of hydrodynamics and 
cavity flow simulations \cite{LFPE-4}.
The lattice Fokker-Planck equation 
is applied to the study of 
electro-rheological transport of 
one-dimensional charged fluid \cite{LFPE-5}
and it is used in phase-space description of 
inertial polymer dynamics \cite{LFPE-6}.
All these lattice Fokker-Planck equations are based on
the lattice Boltzmann discretization approach. 

In this paper, we propose a lattice equation 
for probability density of particle 
in unbounded homogeneous three-dimensional lattice 
with long-range drift and diffusion 
to ${\bf n}$-site from all other ${\bf m}$-sites 
(${\bf m} \ne {\bf n}$).
We prove that continuous limit for
the suggested lattice Fokker-Planck equation gives 
the space-fractional Fokker-Planck equation 
for non-local continuum.
The fractional differential equation for continuum contains 
generalized conjugate Riesz derivatives on non-integer orders.

Continuum mechanics \cite{Sedov} can be considered 
as a continuous limit of lattice dynamics 
\cite{Born,MMW,Bo,Kosevich},
where the length-scales of a continuum element
are much larger than the distances between the lattice particles. 
The first self-consistent derivation of 
the Fokker-Planck equation based on 
the microscopic dynamics for classical and quantum systems 
was obtained by Bogolyubov and Krylov 
\cite{Bogolyubov1,Bogolyubov2}.
Long-range interactions are important
for different problems in statistical mechanics 
\cite{LLI-1,LLI-2,LLI-3},
kinetic theory and nonequilibrium statistical mechanics 
\cite{LLI-4,LLI-5},
theory of non-equilibrium phase transitions
\cite{LLI-6,LLI-7}.
As it was shown in \cite{JPA2006,JMP2006} (see also 
\cite{Chaos2006,CNSNS2006,Laskin} and
\cite{CEJP2013,MOM2014,MPLB2014,ISRN-CMP2014,IJSS2014,ND2014}), 
the continuum equations with fractional derivatives 
can be directly connected to
lattice models with long-range properties. 
A connection between the dynamics of lattice system 
of particles with long-range properties 
and the fractional continuum equations are proved 
by using the transform operation \cite{JPA2006,JMP2006}. 
The papers \cite{JPA2006,JMP2006} deal with 
the one-dimensional lattice models and 
the correspondent one-dimensional continuum equations.
In this paper, we suggest three-dimensional lattice models
for space fractional diffusion processes. 
We propose a general form of three-dimensional lattice 
Fokker-Planck equation, which leads to 
continuum fractional Fokker-Planck equation with 
space derivatives of non-integer orders 
by continuous limit. 
The suggested approach to derive 
the space fractional Fokker-Planck equations 
can serve as a microstructural basis to
describe the spatial-fractional diffusion processes.


\section{Lattice with long-range drift and diffusion}

The lattice is characterized by space periodicity.
In an unbounded lattice we can define three non-coplanar vectors 
${\bf a}_1$, ${\bf a}_2$, ${\bf a}_3$, that 
are the shortest vectors by which a lattice can 
be displaced and be brought back into itself. 
All space lattice sites can be defined by 
the vector ${\bf n} = (n_1,n_2,n_3)$, where $n_i$ are integer. 
For simplification, we consider a lattice with mutually 
perpendicular primitive lattice vectors
${\bf a}_1$, ${\bf a}_2$, ${\bf a}_3$.
We choose directions of the axes of the Cartesian coordinate system coincide with the vector ${\bf a}_i$. Then
${\bf a}_i = a_i \, {\bf e}_i$, 
where $a_i=|{\bf a}_i|$ and ${\bf e}_i$ are 
the basis vectors of the Cartesian coordinate system.
This simplification means that the lattice 
is a primitive orthorhombic Bravais lattice 
with long-range drift and diffusion of particles.

If we choose the coordinate origin at one of the sites, then
the position vector of an arbitrary lattice site 
with ${\bf n} = (n_1,n_2,n_3)$ is written 
${\bf r}({\bf n})=n_1{\bf a}_1+n_2{\bf a}_2+n_3{\bf a}_3$. 
The lattice sites are numbered by ${\bf n}$, so 
that the vector ${\bf n}$ can be considered as a 
number vector of the corresponding particle. 
We assume that the positions of particles coincide with 
the lattice sites ${\bf r}({\bf n})$.
The probability density for lattice site 
will be denoted by $f ({\bf n},t) = f (n_1,n_2,n_3,t)$,
where the site is defined by 
the vector ${\bf n} = (n_1,n_2,n_3)$.
The function $f ({\bf n},t)$ satisfies the conditions
\be
\sum^{+\infty}_{ n_1 = -\infty } \sum^{+\infty}_{ n_2 = -\infty } 
\sum^{+\infty}_{ n_3 = -\infty } f (n_1,n_2,n_3,t) = 1 ,
\quad f (n_1,n_2,n_3,t) \ge 0 
\ee
for all $t \in \mathbb{R}$.

The equation for probability density of particle in 
unbounded homogeneous lattice is
\be \label{LattEq-1}
\frac{\partial f ({\bf n},t)}{\partial t} =
- \sum^3_{i=1} \sum_{ m_i \ne n_i } 
g_i \, K^{i}_{\alpha_i} ({\bf n} - {\bf m}) \, f ({\bf m} ,t) ,
+ \sum^3_{i,j=1} \sum_{ m_i \ne n_i } \sum_{ m_j \ne n_j } 
g_{ij} \, K^{ij}_{\alpha_i,\beta_j} ({\bf n} - {\bf m}) \, 
f ({\bf m} ,t) ,
\ee
where $f({\bf n},t)$ is the probability density function
to find the test particle at site ${\bf n}$ at time $t$.
The italics $i,j \in \{1;2;3\}$ are the coordinate indices, 
$g_{i}$ and $g_{ij}$ are lattice coupling constants. 
The coefficients $K^{i}_{\alpha_i}({\bf n} - {\bf m} )$ and
$K^{ij}_{\alpha_i,\beta_j}({\bf n} - {\bf m} )$ describe
the particle drift and diffusion on the lattice,
and it can be called the drift and diffusion kernels
for lattice step length ${\bf n} - {\bf m}$. 
These kernels describe the long-range drift and 
diffusion to ${\bf n}$-site from all other ${\bf m}$-sites. 
The parameters $\alpha_i$ and $\beta_j$ 
in the kernels are positive real numbers 
that characterize how quickly the intensity of 
the drift and diffusion processes in the lattice
decrease with increasing the value ${\bf n} - {\bf m}$.
These parameters also can be considered as degrees 
of the power law of lattice spatial dispersion 
\cite{CEJP2013,ISRN-CMP2014} that is described 
by non-integer power of the wave vector components.

Equation (\ref{LattEq-1}) describes fractional
diffusion processes on the physical lattices,
where long-range jumps can be realized.
The L\'evy motion (flights) for these lattices
can be described by the lattice 
Fokker-Planck equation (\ref{LattEq-1}),
which is considered
as a lattice analog of the fractional diffusion
processes with the Poissonian waiting time and 
the L\'evy distribution for the jump length \cite{MK2000}.

For simplification, we consider the kernels in the form
\be
K^{i}_{\alpha_i}({\bf n}-{\bf m}) =
K_{\alpha_i}(n_i-m_i) , \quad
K^{ij}_{\alpha_i,\beta_j}({\bf n}-{\bf m}) =
K_{\alpha_i}(n_i-m_i) \, K_{\beta_j}(n_j-m_j) ,
\ee
where $i,j=1,2,3$.
The kernels $K_{\alpha_i}(n_i-m_i)$, where $i=1,2,3$, 
describe long-range jumps in the direction ${\bf a}_i$
with lattice step length $n_i-m_i$
in the lattice.
The correspondent terms with kernels $K_{\alpha_i}(n_i-m_i)$ 
can be considered as lattice analogs of 
fractional derivatives of order $\alpha_i$
with respect to coordinate $x_i=({\bf r},{\bf a}_i)$.
We will consider even and odd types of 
the kernels $K_{\alpha_i}(n_i-m_i)$, $i=1,2,3$, 
that will be denoted by $K^{+}_{\alpha_i}(n_i-m_i)$ 
and $K^{-}_{\alpha_i}(n_i-m_i)$ respectively.


We assume that the kernels $K^{\pm}_{\alpha}(n)$ 
satisfy the following conditions: \\
1) The kernels $K^{\pm}_{\alpha}(n)$ are real-valued functions 
of integer variable $n \in \mathbb{Z}$. 
The kernels $K^{+}_{\alpha}(n)$ and $K^{-}_{\alpha}(n)$ 
are even and odd functions such that
\be
K^{+}_{\alpha}(-n)=+K^{+}_{\alpha} (n) , \quad K^{-}_{\alpha}(-n)=-K^{-}_{\alpha} (n) 
\ee
hold for all $n \in \mathbb{Z}$. \\
2) The kernels $K^{\pm}_{\alpha}(n)$ belong to the Hilbert space 
of square-summable sequences, 
\be \label{Jnm-1b}
\sum^{\infty}_{n=1} |K^{\pm}_{\alpha}(n)|^2 < \infty .
\ee
3) The Fourier series transforms $\hat{K}^{\pm}_{\alpha}(k)$ 
of the kernels $K^{\pm}_{\alpha}(n)$ in the form
\be \label{Jak+}
\hat{K}^{+}_{\alpha}(k)=\sum^{+\infty}_{\substack{n=-\infty \\ n\not=0}} 
e^{-ikn} K^{+}_{\alpha}(n) = 2 \sum^{\infty}_{n=1} K^{+}_{\alpha}(n) \cos(kn) ,
\ee
\be \label{Jak-}
\hat{K}^{-}_{\alpha}(k)=
\sum^{+\infty}_{n=-\infty} 
e^{-ikn} K^{-}_{\alpha}(n) = 
- 2 \, i \, \sum^{\infty}_{n=1} K^{-}_{\alpha}(n) \sin(kn) 
\ee
satisfy the conditions
\be \label{AR+}
\hat{K}^{+}_{\alpha}(k)= |k|^{\alpha} + o(|k|^{\alpha}), 
\quad ( k \rightarrow 0 ),
\ee
and
\be \label{AR-}
\hat{K}^{-}_{\alpha}(k)= 
i \, \operatorname{sgn}(k) \, |k|^{\alpha} + o(|k|^{\alpha}), 
\quad ( k \rightarrow 0 )
\ee
respectively. 
Here the little-o notation $o(|k|^{\alpha})$ 
means the terms that include higher powers of 
$|k|$ than $|k|^{\alpha}$. 
The suggested forms (\ref{AR+}) and (\ref{AR-})
of the Fourier series transforms of 
the kernels $K^{\pm}_{\alpha}(n)$ mean that we consider 
lattices with weak spatial dispersion \cite{CEJP2013}.
The conditions (\ref{AR+}) and (\ref{AR-}) allow us to consider
a wide class of kernels to describe 
the long-range lattice drift and diffusion.


In general, type of dependence of 
the function $\hat{K}^{\pm}_{\alpha}(k)$ 
on the wave-vector $k$ is defined by the type of 
the spatial dispersion in 
the lattice \cite{CEJP2013,ISRN-CMP2014}. 
For a wide class of processes in the lattice,
the wavelength $\lambda$ holds the relation
$a_0 / \lambda \sim k a_0 \ll 1$, where $a_0$ is
the characteristic size of the lattice distance
such that $a_0 = \operatorname{max} 
\{ |{\bf a}_1|, |{\bf a}_2|, |{\bf a}_3| \}$. 
In the case $k a_0 \ll 1$, where $a_0$,
the spatial dispersion of the lattice is weak. 
To describe lattices with such property 
it is enough to know the dependence of 
the function $\hat{K}^{\pm}_{\alpha}(k)$ only 
for small values $k$, and we can replace 
this function by the Taylor's polynomial series.
The weak spatial dispersion of the lattices with 
power-law type of spatial dispersion cannot be described 
by the usual Taylor approximation.
In this case, we should use a fractional Taylor 
series \cite{CEJP2013,ISRN-CMP2014}.
The fractional Taylor series is more adequate 
for approximation of non-integer power-law functions. 
For example, the usual Taylor series for the power-law function
$\hat{K}^{+}_{\alpha}(k) = a_{\alpha} \, k^{\alpha}$
has the infinite by many terms for non-integer $\alpha$.
The fractional Taylor series of order $\alpha$ 
has a finite number of terms for this function, and
the fractional Taylor's approximation is exact.
We can use the fractional Taylor's series 
in the Riemann-Liouville form 
(see Chapter 1. Section 2.6 \cite{SKM})
that can be represented as
\be \label{Tay-RL}
\hat{K}^{+}_{\alpha}(k_j) = 
b_j (\alpha) \, |k_j|^{\alpha} + o(|k_j|^{\alpha}) ,
\ee
where 
\be
b_j (\alpha) = \frac{(\, _0^{RL}D^{\alpha}_{k_j} \hat{K}^{+}_{\alpha} )(0)}{\Gamma (\alpha +1)} ,
\ee
and $\, _0^CD^{\alpha}_{k}$ is 
the Riemann-Liouville fractional derivative 
\cite{KST} of order $0<\alpha <1$ with respect to $k$.
This derivative is defined by
\be
(\, _0^{RL}D^{\alpha}_{k} \hat{K}^{+}_{\alpha})(k) = 
\left( \frac{d}{dk}\right)^n \, \left( \, _0I^{n-\alpha} _{k} \,
\hat{K}^{+}_{\alpha} \right)(k) ,
\ee
where $\, _0I^{\alpha}_k$ is the left-sided Riemann-Liouville 
fractional integral of order $\alpha >0$ 
with respect to $k$ of the form
\be \label{RLI}
( \, _0I^{\alpha}_{k} \, \hat{K}^{+}_{\alpha} )(k) = 
\frac{1}{\Gamma(\alpha )} 
\int^k_0 \frac{ \hat{K}^{+}_{\alpha} (k^{\prime}) \, dk^{\prime}}{(k-k^{\prime})^{1-\alpha} } , \quad (k>0) .
\ee
Using the approximation (\ref{Tay-RL}), 
we neglect a frequency dispersion for simplification, i.e.,
the parameters $b_j (\alpha)$
do not depend on the frequency $\omega$.
In suggested lattice models, we define the kernels 
such that the constants $g_i$ and $g_{ij}$
include the factor $b_j(\alpha)$, and as a result
the conditions (\ref{AR+}) and (\ref{AR-}) hold.


For simplification, we can consider the lattice kernels 
that are defined by the explicit expressions in the form
\be 
\hat{K}^{+}_{\alpha}(k) = |k|^{\alpha} , \quad 
\hat{K}^{-}_{\alpha}(k) = 
i \operatorname{sgn}(k) \, |k|^{\alpha} .
\ee
In this case, the inverse relations to the definitions of 
$\hat{K}^{\pm}_{\alpha}(k)$ by equations (\ref{Jak+}) 
and (\ref{Jak-}) have the forms
\be \label{InverseK}
K^{+}_{\alpha}(n) = \frac{1}{\pi} 
\int^{\pi}_0 k^{\alpha} \, \cos(n \, k) \, dk , \quad 
K^{-}_{\alpha}(n) = - \frac{1}{\pi} 
\int^{\pi}_0 k^{\alpha} \, \sin(n \, k) \, dk . 
\ee
For non-integer real values of the parameter $\alpha $, 
the expressions for the kernels $K^{\pm}_{\alpha}(n-m)$ are 
\be \label{Kn1+}
K^{+}_{\alpha}(n-m) = \frac{\pi^{\alpha}}{\alpha +1} \, _1F_2 \left(\frac{\alpha +1}{2};
\frac{1}{2},\frac{\alpha +3}{2};-\frac{\pi^2\, (n-m)^2}{4} \right) ,
\quad \alpha > -1 ,
\ee
\be \label{Kn1-}
K^{-}_{\alpha}(n-m) = -\frac{\pi^{\alpha+1} \, (n-m)}{\alpha +2} \, _1F_2 
\left(\frac{\alpha +2}{2}; \frac{3}{2},\frac{\alpha +4}{2};
-\frac{\pi^2\, (n-m)^2}{4} \right) , \quad \alpha >-2,
\ee
where $\, _1F_2$ is the Gauss hypergeometric function 
(see Chapter II in \cite{Erdelyi}). 
Note that expressions can be used not only for $\alpha>0$,
but also for some negative values of $\alpha$.

To visualize the properties of the kernels
(\ref{Kn1+}) and (\ref{Kn1-}), 
we give the plots of the functions
\be \label{Knf1+}
F_{+}(x,y) = \frac{\pi^y}{y+1} \, _1F_2 \left(\frac{y+1}{2};
\frac{1}{2},\frac{y+3}{2};-\frac{\pi^2\, x^2}{4} \right) ,
\ee
\be \label{Knf1-}
F_{-}(x,y) = -\frac{\pi^{y+1} \, x}{y+2} \, _1F_2 \left(\frac{y+2}{2};
\frac{3}{2},\frac{y+4}{2};-\frac{\pi^2\, x^2}{4} \right) ,
\ee
where
\be
K^{\pm}_{\alpha}(n-m)= F_{\pm}(n-m,\alpha ) .
\ee
We present the plots of the function (\ref{Knf1+}) 
by Figures 1, 3, 5, and
the plots of (\ref{Knf1-}) by Figures 2, 4, 6
for the same ranges of $x$ and $y>0$.

Let us note some qualitative properties 
that can be seen from Figures 1-6.
We should note that the functions (\ref{Knf1+}) and (\ref{Knf1-})
represent the kernels with (\ref{Kn1+}) and (\ref{Kn1-}) 
that describe the long-range drift and diffusion to $n$-site 
from all other $m$-sites, where $m \in \mathbb{N}$.
Oscillations tell us that the inflow and outflow of probability
periodically change each other,
when the distance $x=n-m$ between sites increases.
The negative values of $F_{\pm}(x,\alpha )$ 
can be interpreted as the probability flux 
from the site, and
the positive values of $F_{\pm}(x,\alpha )$ 
can be interpreted as the flux to the site.
Maximums and minimums of $F_{\pm}(x,\alpha )$ 
characterize an amplitude of oscillation of 
the probability flux from the site and into the site.
The amplitudes as functions of the parameter $\alpha $ 
are increasing functions for a fixed value $x=n$.
Plot of the functions (\ref{Knf1+}) and (\ref{Knf1-}) 
with $\alpha =1.5$ and $\alpha =1$ for the range $x\in[0,7]$
are presented by Figures 7 and 8, where 
the graphics of functions with $\alpha =1.5$ 
have larger amplitudes than 
the graphics of the functions with $\alpha =1$.
The amplitudes as functions of the values $n$ 
are decreasing functions for a fixed value $\alpha $, 
and this decreasing has a power-law form.
During the transition from non-local 
to local case for the functions $F_{\pm}(x,\alpha )$, 
a sharp jump does not occur.
We can only state that the fractional power-law decreasing 
is transformed into the decreasing of 
integer power form.

It should be noted that 
the kernels $K^{+}_{\alpha}(n)$ give the local operators
for continuum limit for even $\alpha$ only,
and $K^{-}_{\alpha}(n)$ give the local operators 
for odd $\alpha$ only.
The kernels $K^{\pm}_{\alpha}(n)$ for integer values of $\alpha$ 
(see also Sec.2.5.3.5 in \cite{Prudnikov}) 
can be represented by the equations
\be \label{KsnGen+}
K^{+}_{\alpha}(n) = \sum^{[(\alpha -1)/2]}_{k=0}
\frac{(-1)^{n+k} \, \alpha ! \, \pi^{\alpha-2k-2}}{(\alpha -2n-1)!} \, 
\frac{1}{n^{2k+2}} + 
\frac{(-1)^{[(\alpha +1)/2]} \, \alpha ! \, ( 2[(\alpha +1)/2]-\alpha ) }{
\pi \, n^{\alpha+1}} ,
\ee
and
\be \label{KsnGen-}
K^{-}_{\alpha}(n) = - \sum^{[\alpha /2]}_{k=0}
\frac{(-1)^{n+k+1} \, \alpha ! \, \pi^{\alpha-2k-1}}{(\alpha -2n)!} \, 
\frac{1}{n^{2k+2}} -
\frac{(-1)^{[\alpha /2]} \, \alpha ! \, ( 2[\alpha /2]-\alpha +1 ) }{
\pi \, n^{\alpha+1}} ,
\ee
where $[z]$ is the integer part of the value $z$,
and $2[(\alpha +1)/2]-\alpha =1$ for odd $n$, and 
$2[(\alpha +1)/2]-\alpha =0$ for even $n$.
We can give examples of kernel with some integer $\alpha $. 
Using equation (\ref{KsnGen+}) or 
direct integration (\ref{InverseK}) for $\alpha \in \{1;2;3\}$, 
we give $K^{+}_{\alpha}(n)$ in the form 
\be \label{2-4}
K^{+}_1(n) = - \frac{1-(-1)^n}{\pi \, n^2} , \quad
K^{+}_2(n) = \frac{2 \, (-1)^n}{n^2} , \quad
K^{+}_3(n) = \frac{3 \, \pi \, (-1)^n}{n^2} +
\frac{6 \, (1-(-1)^n)}{\pi \, n^4} , 
\ee
\be \label{1-3}
K^{-}_1(n) = \frac{(-1)^n}{n} , \quad
K^{-}_2(n) = \frac{(-1)^n\, \pi}{n} +
\frac{2(1-(-1)^n)}{\pi \, n^3}, \quad
K^{-}_3(n) = \frac{(-1)^n\, \pi^2}{n} - 
\frac{6 \, (-1)^n}{n^3} ,
\ee
where $(1-(-1)^n)=2$ for odd $n$, and 
$((-1)^n-1)=0$ for even $n$.




\begin{figure}[H]
\begin{minipage}[h]{0.47\linewidth}
\resizebox{12cm}{!}{\includegraphics[angle=-90]{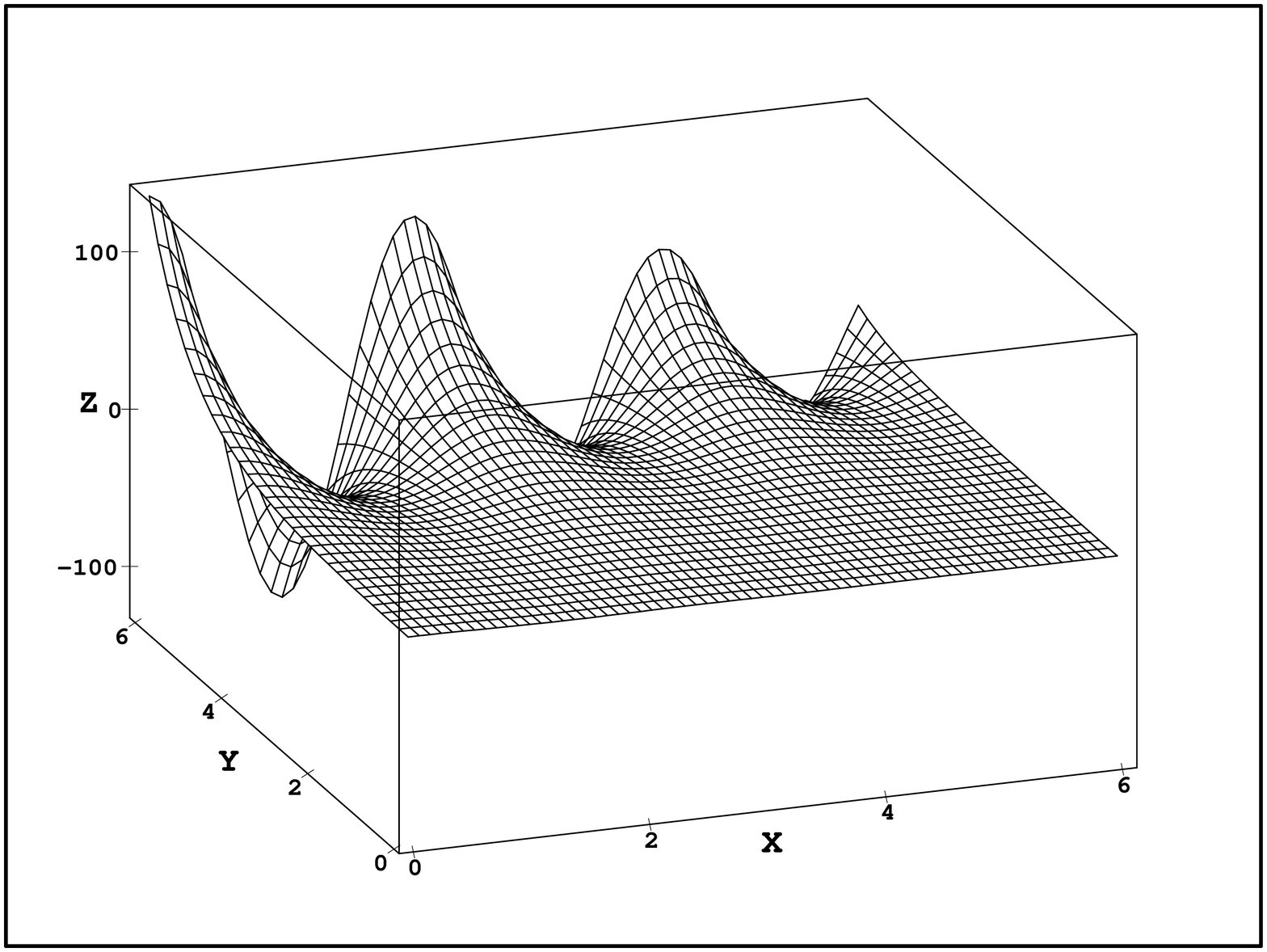} } 
\end{minipage}
\caption{Plot of the function $F_{+}(x,y)$ (\ref{Knf1+}) 
for the range $x\in[0,6]$ and $y=\alpha \in[0,6]$.} 
\label{Plot1.1}
\end{figure}



\begin{figure}[H]
\begin{minipage}[h]{0.47\linewidth}
\resizebox{12cm}{!}{\includegraphics[angle=-90]{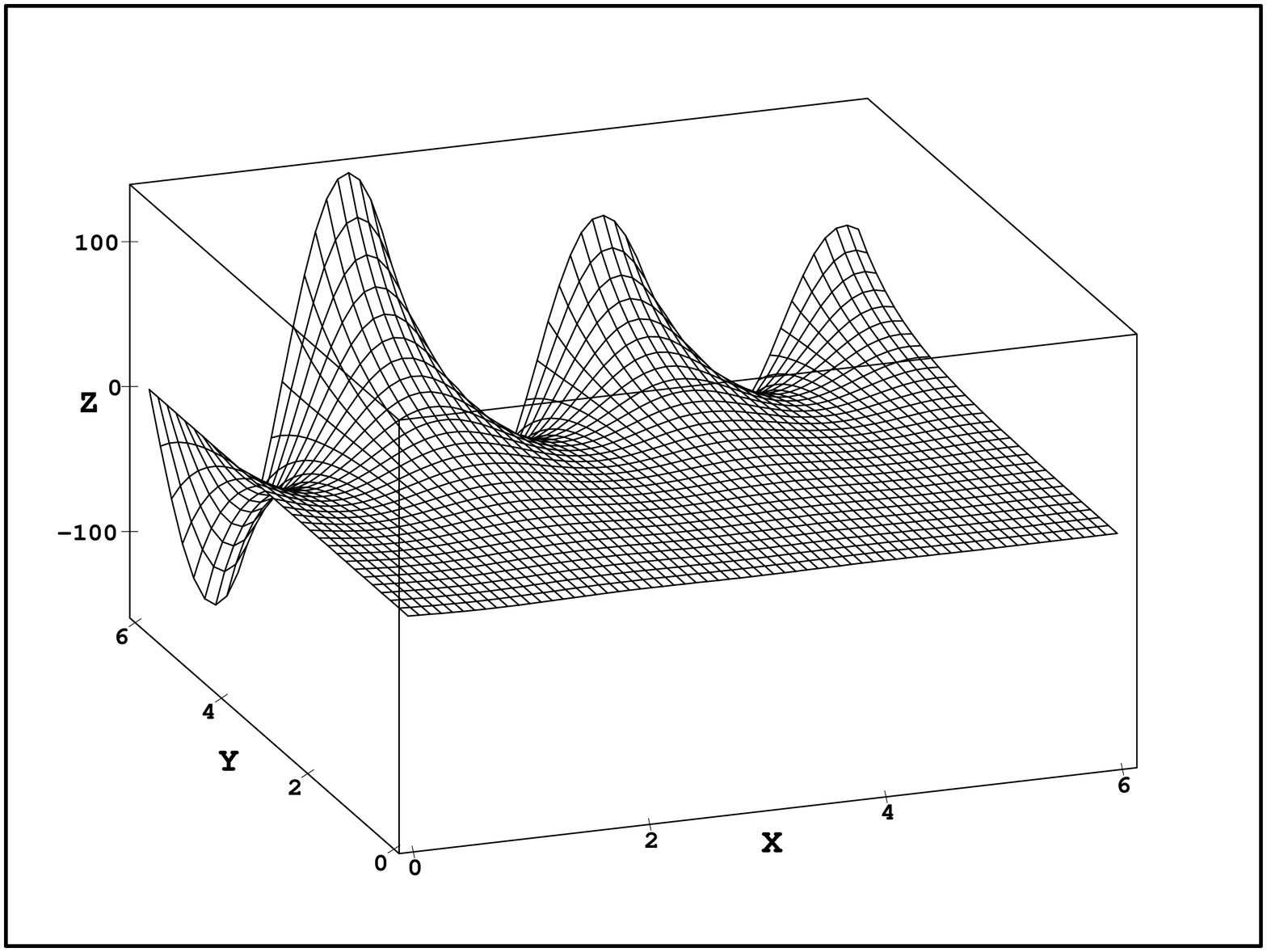} } 
\end{minipage}
\caption{Plot of the function $F_{-}(x,y)$ (\ref{Knf1-})
for the range $x\in[0,6]$ and $y=\alpha \in[0,6]$.} 
\label{Plot2.1}
\end{figure}


\begin{figure}[H]
\begin{minipage}[h]{0.47\linewidth}
\resizebox{12cm}{!}{\includegraphics[angle=-90]{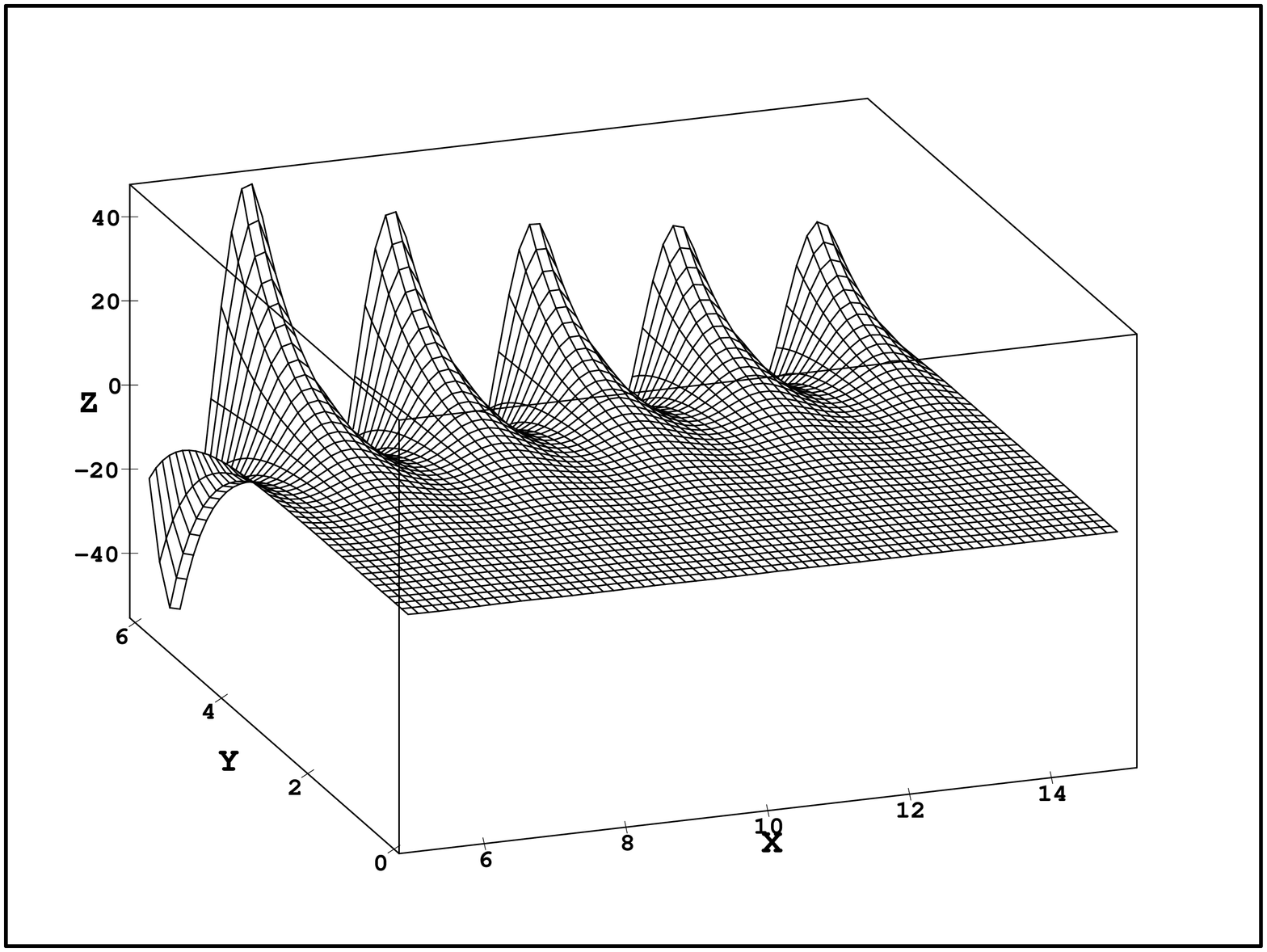} } 
\end{minipage}
\caption{Plot of the function $F_{+}(x,y)$ (\ref{Knf1+})
for the range $x\in[5,15]$ and $y=\alpha \in[0,6]$.} 
\label{Plot1.2}
\end{figure}


\begin{figure}[H]
\begin{minipage}[h]{0.47\linewidth}
\resizebox{12cm}{!}{\includegraphics[angle=-90]{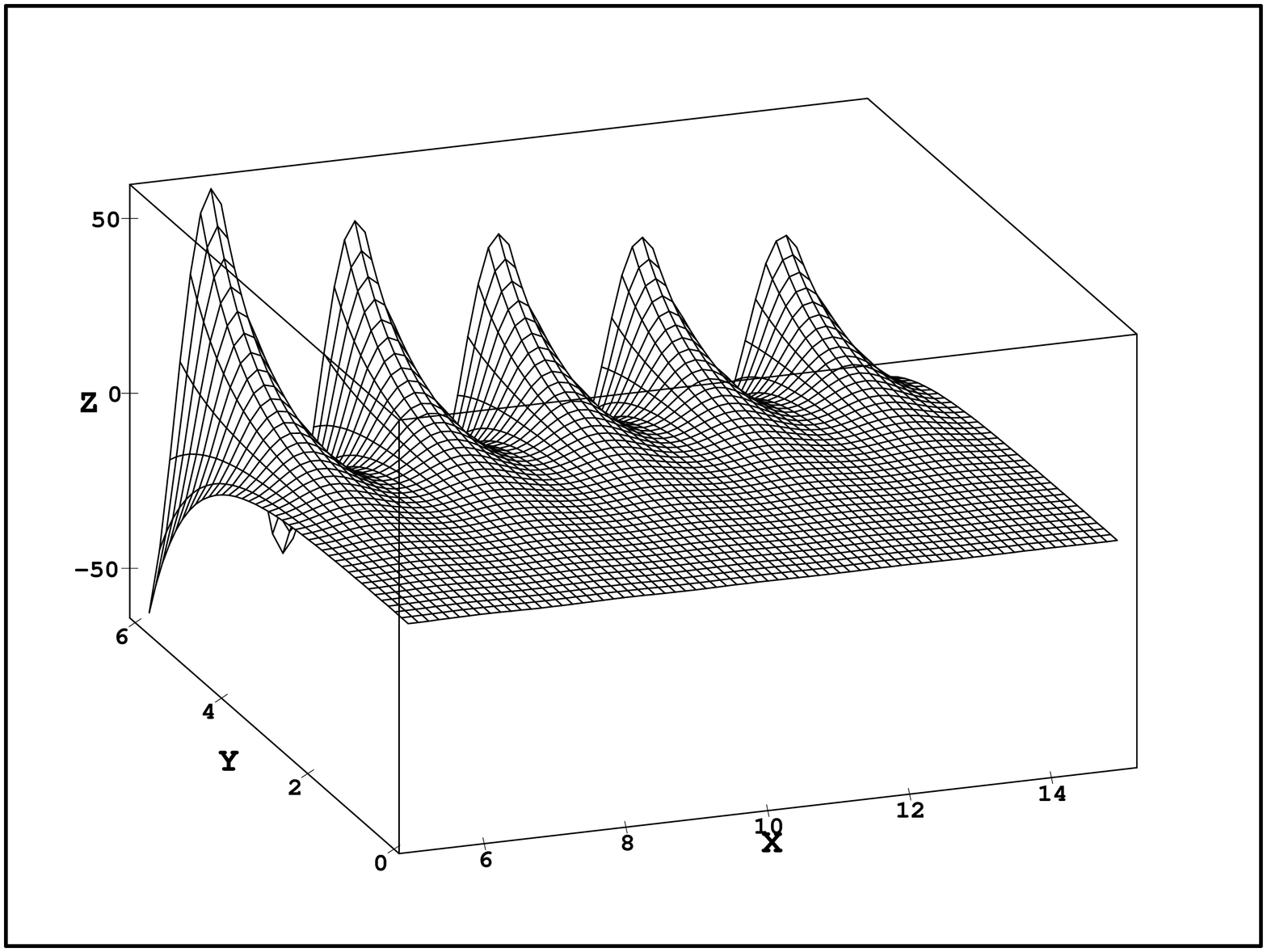} } 
\end{minipage}
\caption{Plot of the function $F_{-}(x,y)$ (\ref{Knf1-})
for the range $x\in[5,15]$ and $y=\alpha \in[0,6]$.} 
\label{Plot2.2}
\end{figure}


\begin{figure}[H]
\begin{minipage}[h]{0.47\linewidth}
\resizebox{12cm}{!}{\includegraphics[angle=-90]{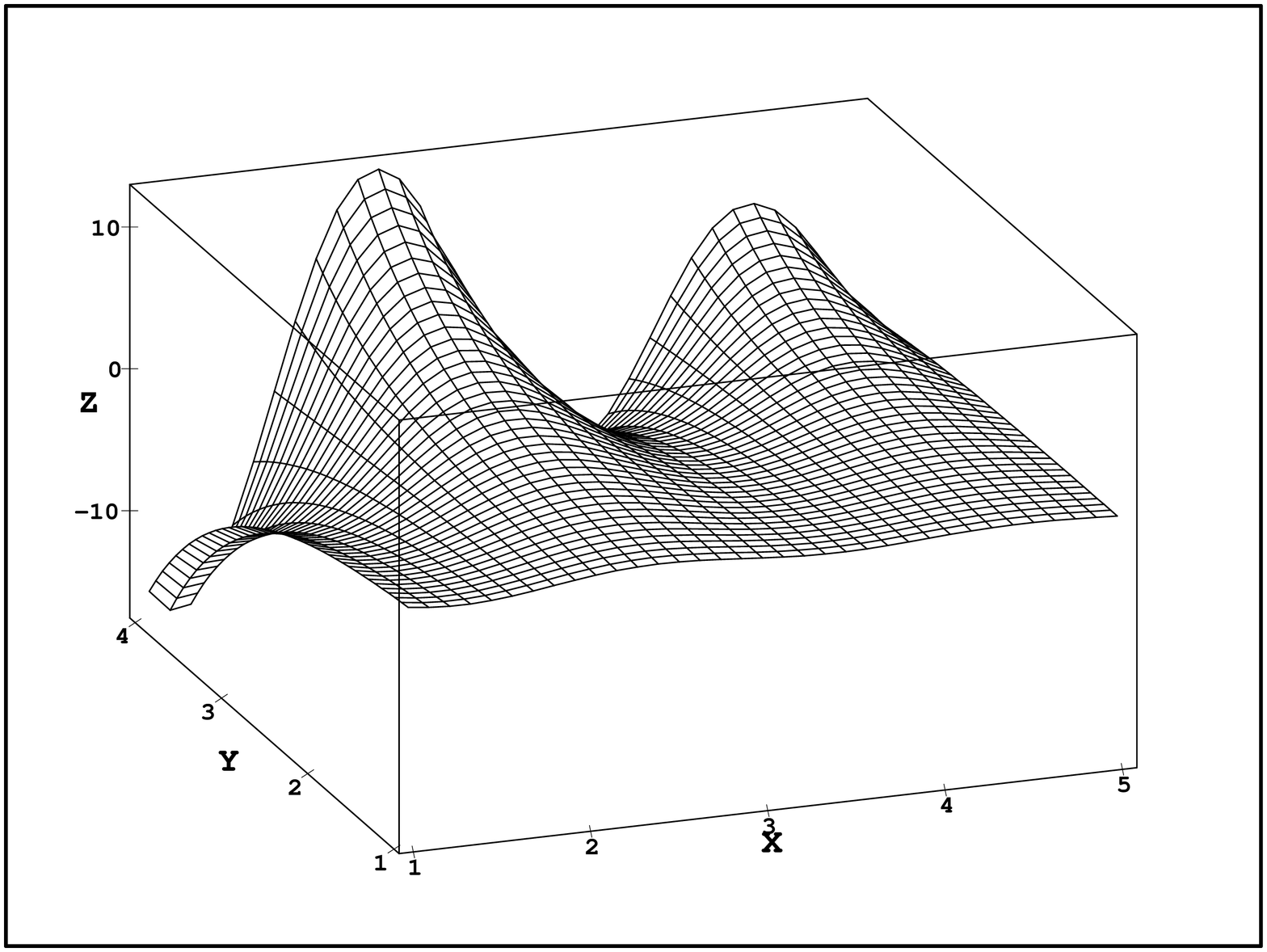} } 
\end{minipage}
\caption{Plot of the function $F_{+}(x,y)$ (\ref{Knf1+})
for the range $x\in[1,5]$ and $y=\alpha \in[1,4]$.} 
\label{Plot1.3}
\end{figure}


\begin{figure}[H]
\begin{minipage}[h]{0.47\linewidth}
\resizebox{12cm}{!}{\includegraphics[angle=-90]{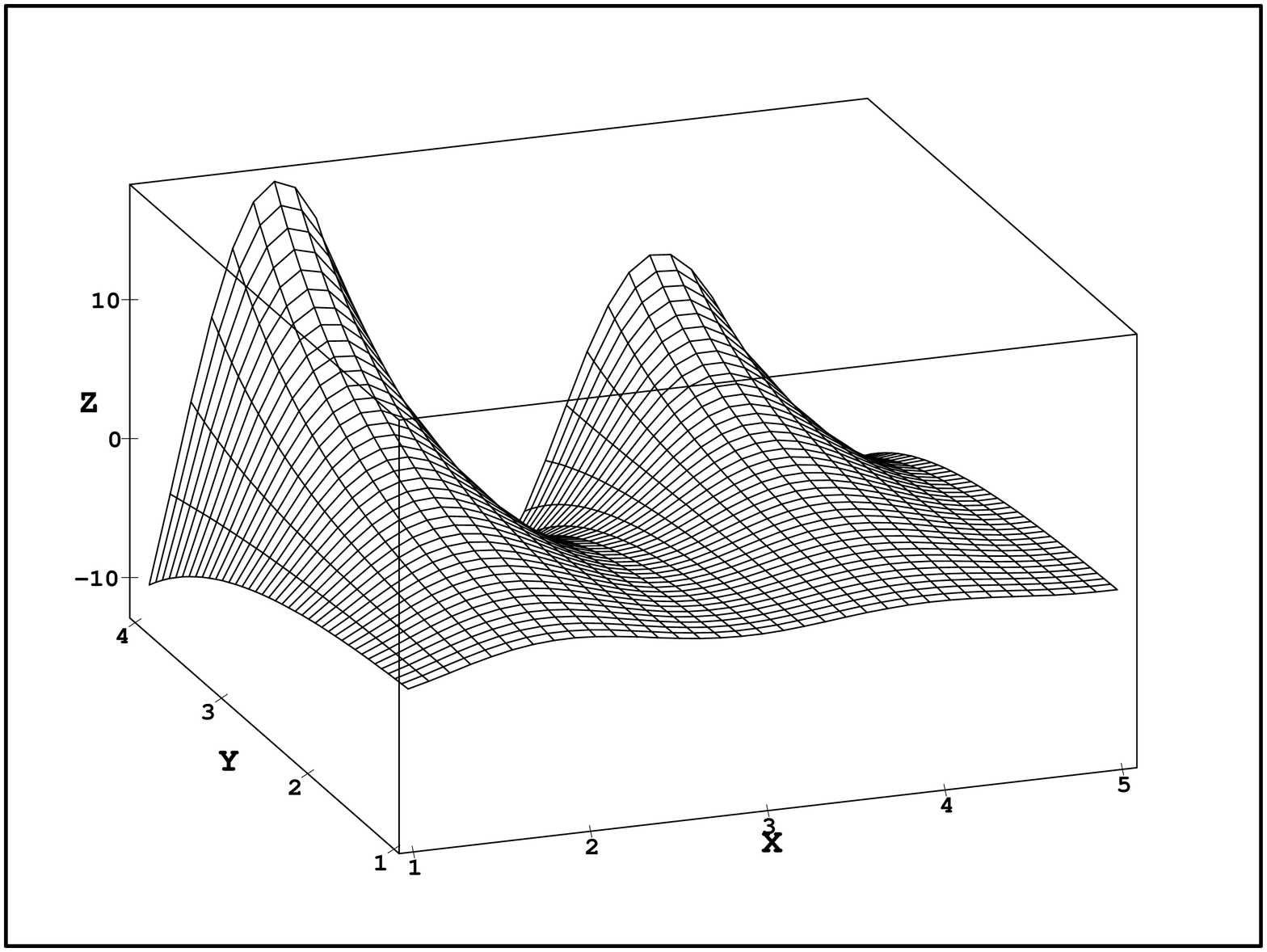} } 
\end{minipage}
\caption{Plot of the function $F_{-}(x,y)$ (\ref{Knf1-})
for the range $x\in[1,5]$ and $y=\alpha \in[1,4]$.} 
\label{Plot2.3}
\end{figure}



\begin{figure}[H]
\begin{minipage}[h]{0.47\linewidth}
\resizebox{12cm}{!}{\includegraphics[angle=-90]{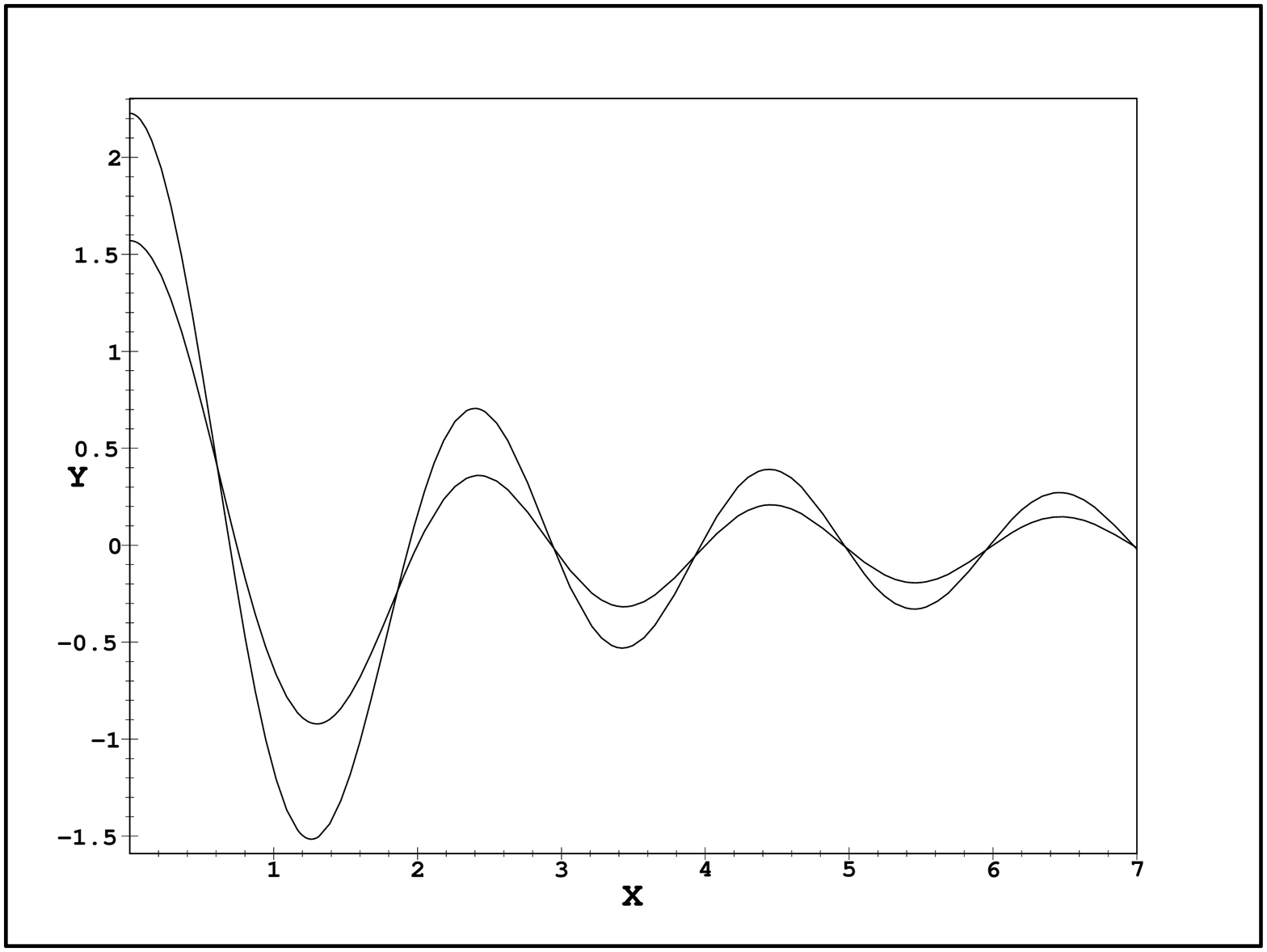} } 
\end{minipage}
\caption{Plot of the function $F_{+}(x)$ (\ref{Knf1+})
with $\alpha =1.5$ and $\alpha =1$ for the range $x\in[0,7]$.} 
\label{Plot3.3}
\end{figure}


\begin{figure}[H]
\begin{minipage}[h]{0.47\linewidth}
\resizebox{12cm}{!}{\includegraphics[angle=-90]{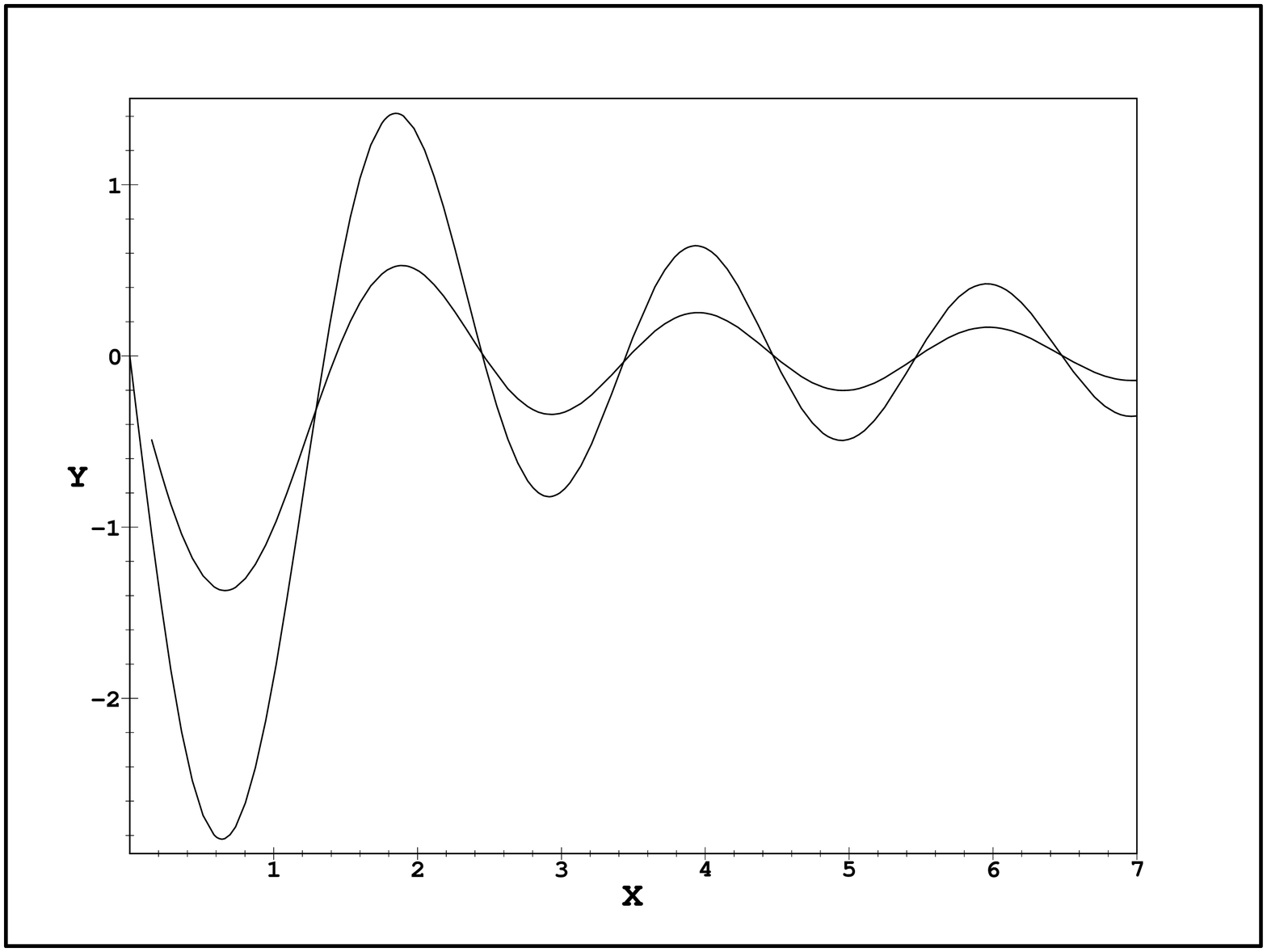} } 
\end{minipage}
\caption{Plot of the function $F_{-}(x)$ (\ref{Knf1-})
with $\alpha =1.5$ and $\alpha =1$ for the range $x\in[0,7]$.} 
\label{Plot3.4}
\end{figure}


For simplification of the form of lattice equation,
we use the lattice operators 
$\mathbb{K}^{\pm} \left[ \alpha\atop i \right]$
such that the action of these operators on 
the lattice probability density $f({\bf m},t)$ is
\be \label{Ds1}
\mathbb{K}^{\pm} \left[ \alpha\atop i \right] f ({\bf m},t) = 
\sum_{\substack{m_i=-\infty \\ m_i \ne n_i}}^{+\infty} \; 
K^{\pm}_{\alpha} (n_i-m_i) \; f ({\bf m},t) ,
\quad (i=1,2,3) .
\ee
The values $i=1,2,3$ specify one of 
the three variables $n_1$, $n_2$, $n_3$ of the lattice site 
that are similar to $x_i$ of the space $\mathbb{R}^3$.
If $\alpha_i=1$, then $\mathbb{K}^{+} $
is a nonlocal operator, and
if $\alpha_i=2$, then $\mathbb{K}^{-}$ are 
nonlocal operators also. 
Note that the operators 
$\mathbb{K}^{+} \left[ \alpha_i \atop i \right]$
for odd integer values of $\alpha_i$ and
$\mathbb{K}^{-} \left[ \alpha_i \atop i \right]$
for even integer values of $\alpha_i$ are nonlocal.
For example, the operators
$\mathbb{K}^{+} \left[ 1 \atop i \right]$
and $\mathbb{K}^{-} \left[ 2 \atop i \right]$
cannot be considered as a local operators of integer orders.

We also can consider combinations of the lattice operators
\be \label{Dsss1a}
\mathbb{K}^{\pm,\pm} \left[ \alpha_i \, \beta_j \atop i \ j \right] =
\mathbb{K}^{\pm} \left[ \alpha_i \atop i \right] \,
\mathbb{K}^{\pm} \left[ \beta_j \atop j \right] ,
\ee
where $i$, $j$ take values from the set $\{1;2;3\}$.
The action of the operator (\ref{Dsss1a}) on 
the lattice probability density $f({\bf m},t)$ is
\be \label{Dsss1b}
\mathbb{K}^{\pm,\pm} \left[ \alpha_i \, \beta_j \atop i \ j \right] 
f ({\bf m},t) = 
\sum_{\substack{m_i=-\infty \\ m_i \ne n_i}}^{+\infty} \; 
\sum_{\substack{m_j=-\infty \\ m_j \ne n_j}}^{+\infty} \; 
K^{\pm}_{\alpha_i} (n_i-m_i) \; K^{\pm}_{\beta_j} (n_j-m_j) \; 
f ({\bf m},t) .
\ee
This is the mixed lattice operators.

Using the lattice operators (\ref{Ds1}) and (\ref{Dsss1b}), 
the equation for probability density (\ref{LattEq-1})
takes the form
\be \label{LattEq-2}
\frac{\partial f ({\bf n},t)}{\partial t} =
- \sum^3_{i=1} g_i \,
\mathbb{K}^{\pm} \left[ \alpha_i \atop i \right]
\, f ({\bf m} ,t) ,
+ \sum^3_{i,j=1} g_{ij} \, \mathbb{K}^{\pm,\pm} 
\left[ \alpha_i \, \beta_j \atop i \ j \right] 
 \, f ({\bf m} ,t) .
\ee
This is the three-dimensional 
lattice Fokker-Planck equation in the operator form 
to describe fractional diffusion and drift
with the lattice jump length $({\bf n}-{\bf m})$.

To describe the long-range drift and diffusion for 
the lattice with memory, we can use the equation
\be \label{LattEq-2m}
\frac{\partial f ({\bf n},t)}{\partial t} =
- \sum^3_{i=1} g_i \,
\mathbb{K}^{\pm} \left[ \alpha_i \atop i \right]
\, f ({\bf m} ,t) ,
+ \sum^3_{i,j=1} g_{ij} \, \mathbb{K}^{\pm,\pm} 
\left[ \alpha_i \, \beta_j \atop i \ j \right] \,
_0^{RL}D^{1-\gamma}_t \, f ({\bf m} ,t) ,
\ee
where $_0^{RL}D^{1-\gamma}_t $ is the Riemann-Liouville fractional derivative of order $(1-\gamma)$ with respect to time \cite{KST}.
Note that the time-fractional derivatives $_0^{RL}D^{1-\gamma}_t$ 
is present only in the diffusion term. 
This fractional derivative describes the long-term memory
of power-law type. Equation (\ref{LattEq-2m}) describes 
anomalous diffusion processes with the waiting time $t$
and the lattice jump length $({\bf n}-{\bf m})$.


We can consider the time-fractional derivatives 
$_0^{RL}D^{1-\gamma}_t$ in the first and second terms
of the right side of 
the lattice Fokker-Planck equation (\ref{LattEq-2}).
In this case the time-fractional lattice 
Fokker-Planck equation has the form
\be \label{LattEq-2mm}
\frac{\partial f ({\bf n},t)}{\partial t} = \,
_0^{RL}D^{1-\gamma}_t \, 
{\cal L}^{\alpha, \beta}_{LFP} \, f ({\bf m} ,t) ,
\ee
where ${\cal L}^{\alpha, \beta}_{LFP}$ is the lattice
Fokker-Planck operator
\be
{\cal L}^{\alpha, \beta}_{LFP} =
- \sum^3_{i=1} g_i \,
\mathbb{K}^{\pm} \left[ \alpha_i \atop i \right]
+ \sum^3_{i,j=1} g_{ij} \, \mathbb{K}^{\pm,\pm} 
\left[ \alpha_i \, \beta_j \atop i \ j \right] .
\ee
Equation (\ref{LattEq-2mm}) describes long-range diffusion 
and drift with power-law memory on orthorhombic Bravais lattices.


\section{Continuum limit for lattice equations} 

\subsection{Continuum limit for lattice probability density}

In order to transform a lattice probability 
density $f ({\bf n},t)$
into a probability density $f ({\bf r},t)$ of continuum, 
we use approach suggested in \cite{JMP2006,JPA2006}.
We propose to consider $f ({\bf n},t)$ as 
Fourier series coefficients
of some function $\hat{f}({\bf k},t)$ for 
$k_j \in [-k_{j0}/2, k_{j0}/2]$, where $k_{j0}=2 \pi/ a_j$.
Then we use the continuous limit ${\bf k}_0 \to \infty$ 
to obtain $\tilde{f}({\bf k},t)$,
and finally we apply the inverse Fourier integral transformation to obtain the probability density $f ({\bf r},t)$. 
For clarity, we have presented the set of transformations of
the probability density by Figure 9.


\begin{picture}(350,250)
\multiput(50,200)(150,0){3}{\oval(90,60)}
\multiput(50,100)(300,0){2}{\oval(128,60)}
\multiput(50,0)(150,0){3}{\oval(90,60)}
\put(30,195){$f ({\bf n},t)$}
\put(165,210){\text{From Lattice}}
\put(165,190){\text{to Continuum}}
\put(330,195){$f ({\bf r},t)$}
\put(10,110){\text{Fourier series} }
\put(10,90){\text{transform ${\cal F}_{\Delta}$}}
\put(300,110){\text{Inverse Fourier }}
\put(292,90){\text{integral transform ${\cal F}^{-1}$}}
\put(35,-5){$\hat{f}({\bf k},t)$}
\put(180,7){\text{$\operatorname{Limit}$ }}
\put(180,-14){\text{$a_j \to 0$}}
\put(335,-5){$\tilde{f}({\bf k},t)$}
\put(45,170){\line(0,-1){40}}
\put(55,170){\line(0,-1){40}}
\put(45,70){\line(0,-1){30}}
\put(55,70){\line(0,-1){30}}
\put(345,160){\line(0,-1){30}}
\put(355,160){\line(0,-1){30}}
\put(345,70){\line(0,-1){40}}
\put(355,70){\line(0,-1){40}}
\put(50,30){\line(-1,2){10}}
\put(50,30){\line(1,2){10}}
\put(350,170){\line(-1,-2){10}}
\put(350,170){\line(1,-2){10}}
\multiput(95,5)(0,200){2}{\line(1,0){60}}
\multiput(95,-5)(0,200){2}{\line(1,0){60}}
\multiput(245,5)(0,200){2}{\line(1,0){50}}
\multiput(245,-5)(0,200){2}{\line(1,0){50}}
\multiput(305,0)(0,200){2}{\line(-2,1){20}}
\multiput(305,0)(0,200){2}{\line(-2,-1){20}}
\put(0,-60){\text{Figure 9: Transformation 
of lattice probability density
to continuum probability density.}}
\end{picture}
\vskip 30mm

The transformation a lattice probability density into 
a continuum probability density
is realized by a sequence of the following three steps: 

The first step is the Fourier series transform 
${\cal F}_{\Delta}: \quad f ({\bf n},t) \to 
{\cal F}_{\Delta} \{ f ({\bf n},t)\} = \hat{f}({\bf k},t) $ 
that is defined by
\be \label{ukt}
\hat{f}({\bf k},t) = 
\sum_{n_1,n_2,n_3=-\infty}^{+\infty} \; f ({\bf n},t) 
\; e^{-i ({\bf k}, {\bf r}({\bf n})) } =
{\cal F}_{\Delta} \{f ({\bf n},t)\} ,
\ee
where the inverse transformation is
\be \label{un} 
f ({\bf n},t) = \left(\prod^3_{j=1} 
\frac{1}{k_{j0}}\right)
\int_{-k_{10}/2}^{+k_{10}/2} dk_1  . . . 
\int_{-k_{30}/2}^{+k_{30}/2} dk_3
\ \hat{f}({\bf k},t) \; e^{i ({\bf k}, {\bf r}({\bf n}))}= 
{\cal F}^{-1}_{\Delta} \{ \hat{f}({\bf k},t) \} ,
\ee
and ${\bf r}({\bf n}) = \sum^3_{j=1} n_j \, {\bf a}_j $ 
and $k_{j0}=2\pi/a_j$.
We assume that all lattice particles have 
the same inter-particle distance $a_j$
in the direction ${\bf a}_j$ for simplification.

The second step is the passage to the limit 
$a_j \to 0$ ($k_{j0} \to \infty$) denoted by
$\operatorname{Lim}: \quad \hat{f}({\bf k},t) 
\to \operatorname{Lim} \{\hat{f}({\bf k},t)\}
= \tilde{f}({\bf k},t)$. 
The function $\tilde{f}({\bf k},t)$ 
can be derived from $\hat{f}({\bf k},t)$
in the limit $a_i \to 0$.
Note that $\tilde{f}({\bf k},t)$ 
is a Fourier integral transform of the 
probability density $f ({\bf r},t)$,
and $\hat{f}(k,t)$ is a Fourier series transform 
of $f ({\bf n},t)$, where we use 
\[ f ({\bf n},t) = \prod^3_{j=1} \frac{2 \pi}{k_{j0}} f ({\bf r(n)},t)  \] 
considering ${\bf r(n)}=\sum^3_{j=1} n_j {\bf a}_j= 
2 \pi \sum^3_{j=1} n_j /k_{j0} {\bf e}_j \to {\bf r}$.

The third step is the inverse Fourier integral transform 
${\cal F}^{-1}: \quad \tilde{f}({\bf k},t) 
\to {\cal F}^{-1} \{ \tilde{f}({\bf k},t)\}=f ({\bf r},t)$
is defined by 
\be \label{uxt}
f ({\bf r},t) =\frac{1}{(2\pi)^3} 
\int^{+\infty}_{-\infty} d^3 {\bf k} 
\ e^{i \sum^3_{j=1}k_jx_j} 
\tilde{f}({\bf k},t) =
{\cal F}^{-1} \{ \tilde{f}({\bf k},t) \} 
\ee
that corresponds to the transformation
\be \label{ukt2} 
\tilde{f}({\bf k},t)=
\int^{+\infty}_{-\infty} d^3 {\bf r} \ 
e^{-i \sum^3_{j=1} k_jx_j} f ({\bf r},t) = 
{\cal F} \{ f({\bf r},t) \} .
\ee

Note that the Fourier series transform equations 
(\ref{ukt}) and (\ref{un}) 
in the limit $a_j \to 0$ ($k_{j0} \to \infty$) 
give the Fourier integral transform 
equations (\ref{ukt2}) and (\ref{uxt}),
where the sum is replaced by the integral.

The lattice probability density $f ({\bf n},t)$
is transformed by the combination 
${\cal F}^{-1} \circ \, \operatorname{Lim} \circ 
\, {\cal F}_{\Delta}$ 
into a probability density $f ({\bf r},t)$ of continuum, 
\be \label{Limit-1} 
{\cal F}^{-1} \circ \, \operatorname{Lim} \circ 
\, {\cal F}_{\Delta} \Bigl( f ({\bf n},t) \Bigr) =
f ({\bf r},t).
\ee

The combination of the operations 
${\cal F}^{-1}$, $\operatorname{Lim}$, and ${\cal F}_{\Delta}$ 
allows us to map the lattice functions and operators
into functions and operators for continuum.


\subsection{Continuum limit of lattice operators} 


Let us consider transformations of lattice operators 
(\ref{Ds1}) and (\ref{Dsss1a}) into continuum operators.
The transformations 
${\cal F}^{-1} \circ \operatorname{Limit} \circ 
\ {\cal F}_{\Delta}$ map
the lattice operators into the fractional derivatives 
with respect to coordinates.
We represent these transformations by Figure 10.


\begin{picture}(350,250)
\multiput(50,200)(150,0){3}{\oval(90,60)}
\multiput(50,100)(300,0){2}{\oval(120,60)}
\multiput(50,0)(150,0){3}{\oval(90,60)}
\put(30,195){\text{{\large $\mathbb{K}^{\pm} 
\left[ \alpha \atop i \right]$}}}
\put(165,210){\text{From Lattice}}
\put(165,190){\text{to Continuum}}
\put(330,195){\text{{\Large $\frac{\partial^{\alpha,\pm}}{\partial |x_i|^{\alpha}}$}}}
\put(10,110){\text{Fourier series} }
\put(10,90){\text{transform ${\cal F}_{\Delta}$}}
\put(300,110){\text{Inverse Fourier }}
\put(292,90){\text{integral transform ${\cal F}^{-1}$}}
\put(35,-5){\text{$\hat{K}^{\pm}_{\alpha}(k)$}}
\put(180,7){\text{$\operatorname{Limit}$ }}
\put(180,-14){\text{$a_j \to 0$}}
\put(330,-5){\text{$\tilde{K}^{\pm}_{\alpha}(k)$}}
\put(45,170){\line(0,-1){40}}
\put(55,170){\line(0,-1){40}}
\put(45,70){\line(0,-1){30}}
\put(55,70){\line(0,-1){30}}
\put(345,160){\line(0,-1){30}}
\put(355,160){\line(0,-1){30}}
\put(345,70){\line(0,-1){40}}
\put(355,70){\line(0,-1){40}}
\put(50,30){\line(-1,2){10}}
\put(50,30){\line(1,2){10}}
\put(350,170){\line(-1,-2){10}}
\put(350,170){\line(1,-2){10}}
\multiput(95,5)(0,200){2}{\line(1,0){60}}
\multiput(95,-5)(0,200){2}{\line(1,0){60}}
\multiput(245,5)(0,200){2}{\line(1,0){50}}
\multiput(245,-5)(0,200){2}{\line(1,0){50}}
\multiput(305,0)(0,200){2}{\line(-2,1){20}}
\multiput(305,0)(0,200){2}{\line(-2,-1){20}}
\put(0,-60){\text{Figure 10: Transformation 
of lattice operators to
fractional derivatives.}}
\end{picture}
\vskip 30mm


Using the methods suggested in \cite{JMP2006,JPA2006},
we can prove the connection between the lattice operators
and fractional derivatives of non-integer orders
with respect to coordinates. 


The lattice operators (\ref{Ds1}), 
where $K^{\pm}_{\alpha}(n-m)$ are defined by 
(\ref{Kn1+}) and (\ref{Kn1-}), 
are transformed by the combination 
${\cal F}^{-1} \circ \, \operatorname{Lim} \circ 
\, {\cal F}_{\Delta}$ 
into the fractional derivatives of order $\alpha $ 
with respect to coordinate $x_i$ as 
\be \label{Limit-2}
{\cal F}^{-1} \circ \, \operatorname{Lim} \circ 
\, {\cal F}_{\Delta} 
\Bigl( \mathbb{K}^{\pm} \left[ \alpha\atop i \right] \Bigr) =
a^{\alpha}_i \, \frac{\partial^{\alpha,\pm} }{\partial |x_i|^{\alpha} } ,
\ee
where $a_i=|{\bf a}_i|$ are the primitive lattice vectors,
${\partial^{\alpha,+}}/{\partial |x_i|^{\alpha}}$
is the Riesz fractional derivative of order $\alpha >0$
with respect to $x_i$, 
and ${\partial^{\alpha,-}}/{\partial |x_i|^{\alpha}}$
is the generalized conjugate Riesz derivative
of order $\alpha >0$.
The order of the partial derivative
${\partial^{\alpha,\pm}}/{\partial |x_i|^{\alpha} }$
is defined by the order of lattice operator
$\mathbb{K}^{\pm} \left[ \alpha\atop i \right]$
and it can be integer and non-integer.


Using the independence of the site vectors 
of lattice site ${\bf n}_1 = (n_1,0,0)$, 
${\bf n}_2 = (0,n_2,0)$, ${\bf n}_3 = (0,0,n_3)$ 
and the statement (\ref{Limit-2}), we can prove that
the continuum limits for the mixed lattice operators
(\ref{Dsss1a}) have the form
\be \label{Limit-3}
{\cal F}^{-1} \circ \, \operatorname{Lim} \circ 
\, {\cal F}_{\Delta} 
\Bigl( \mathbb{K}^{\pm,\pm} 
\left[ \alpha_i \, \beta_j \atop i \ j \right]\Bigr) = 
a^{\alpha_i}_i \, a^{\alpha_j}_j \, 
\frac{\partial^{\alpha_i,\pm} }{\partial |x_i|^{\alpha_i} } 
\frac{\partial^{\beta_j,\pm} }{\partial |x_j|^{\beta_j} } ,
\ee
As a result, we obtain continuum limits for
the lattice fractional derivatives in the form of 
the fractional derivatives of the Riesz type
with respect to coordinates.


The Riesz fractional derivative of the order $\alpha$ 
is defined \cite{SKM,KST} by the equation
\be \label{Rder} 
\frac{\partial^{\alpha,+} f({\bf r})}{\partial |x_i|^{\alpha} }
= \frac{1}{d_1(m,\alpha)} \int_{\mathbb{R}} 
\frac{1}{|z_i|^{\alpha+1}} (\Delta^m_i f)({\bf z}_i) \, dz_i , 
\quad (0 <\alpha <m) ,
\ee
where $(\Delta^m_i f)({\bf z}_i)$ is a finite difference of
order $m$ of a function $f({\bf r})$ with the vector step 
${\bf z}_i= x_i \, {\bf e}_i \in \mathbb{R}^3$
for the point ${\bf r} \in \mathbb{R}^3$.
The non-centered difference is
\be \label{ncent}
(\Delta^m_{i} f)({\bf z}_i) =\sum^m_{k=0} (-1)^k 
\frac{m!}{k! \, (m-k)!} \, f({\bf r}-k \, {\bf z}_i) , 
\ee
and the centered difference
\be \label{cent}
(\Delta^m_{i} f)({\bf z}_i) =\sum^m_{k=0} (-1)^k 
\frac{m!}{k! \, (m-k)!} \, f({\bf r}-(m/2-k) \, {\bf z}_i) . 
\ee
The constant $d_1(m,\alpha)$ is defined by
\[ d_1(m,\alpha) =\frac{\pi^{3/2} A_m(\alpha)}{2^{\alpha} 
\Gamma(1+\alpha/2) \Gamma((1+\alpha)/2) \sin (\pi \alpha/2)} , \]
where 
\[ A_m(\alpha)=2 \sum^m_{j=0} (-1)^{j-1} \frac{m!}{j!(m-j)!} \, j^{\alpha} \]
in the case of the non-centered difference (\ref{ncent}), and
\[ A_m(\alpha)=2\, \sum^{[m/2]}_{j=0} (-1)^{j-1} 
\frac{m!}{j!(m-j)!} \, (m/2- j)^{\alpha} \]
in the case of the centered difference (\ref{cent}).
The constants $d_1(m,\alpha)$ is different from zero
for all $\alpha >0$ in the case of an even $m$ and centered difference $(\Delta^m_{i} f)$ (see Theorem 26.1 in \cite{SKM}).
In the case of a non-centered difference the constant
$d_1(m,\alpha)$ vanishes if and only if 
$\alpha=1,3,5,...,2[m/2]-1$.
Note that the integral (\ref{Rder})
does not depend on the choice of $m>\alpha$.
The Fourier transform ${\cal F}$ of the Riesz fractional 
derivative is given by 
\be \label{FdelZ}
{\cal F} \left(
\frac{\partial^{\alpha,+} f({\bf r})}{\partial |x_i|^{\alpha} }
\right)({\bf k}) = |k_i|^{\alpha} ({\cal F} f)({\bf k}) . 
\ee
Equation (\ref{FdelZ}) can be considered as a definition
of the Riesz fractional derivative of order $\alpha$.

Using that $(-i)^{2j}=(-1)^{j}$, 
the Riesz derivatives for even $\alpha=2j$ are
\be \label{s=2j}
\frac{\partial^{2j,+} f({\bf r})}{\partial |x_i|^{2j} }
= (-1)^j \, \frac{\partial^{2j} f({\bf r})}{\partial x_i^{2j} } .
\ee
For $\alpha=2$ the Riesz derivative looks like the Laplace operator. 
The fractional derivatives 
${\partial^{\alpha,+}}/{\partial |x_i|^{\alpha}}$
for even orders $\alpha$ are local operators.
Note that the Riesz derivative
${\partial^{1,+}}/{\partial |x_i|^{1}}$
cannot be considered as a derivative of first order
with respect to $|x_i|$.
For $\alpha=1$ it is it looks like the square root of the Laplacian. 
The Riesz derivatives for odd orders $\alpha=2j+1$
are non-local operators that cannot be considered
as usual derivatives ${\partial^{2j+1}}/{\partial x^{2j+1}_i}$.


We also define the new fractional derivatives 
${\partial^{\alpha,-}}/{\partial |x_i|^{\alpha}}$ 
by the equation 
\be \label{CRFD}
\frac{\partial^{\alpha,-}}{\partial |x_i|^{\alpha}} =
\left\{
\aligned
\frac{\partial}{\partial x_i} \, {\bf I}^{1-\alpha}_i & 
\qquad 0<\alpha<1 , \\
\frac{\partial}{\partial x_i} & 
\qquad \alpha=1 , \\
\frac{\partial}{\partial x_i} \, 
\frac{\partial^{\alpha-1,+}}{\partial |x_i|^{\alpha-1}} &
\qquad \alpha>1 , 
\endaligned
\right.
\ee
where $\partial/\partial x_i$ is the usual derivative
of first order with respect to coordinate $x_i$,
and ${\bf I}^{1-\alpha}_i$ is the Riesz potential of order $(1-\alpha)$ (see Appendix) with respect to $x_i$,
\be \label{RFP-i}
{\bf I}^{1-\alpha}_i f({\bf r}) = 
\int_{\mathbb{R}^1} R_{1-\alpha}(x_i-z_i) \
f({\bf r}+ (z_i-x_i){\bf e}_i) dz_i, \quad (0<\alpha <1) ,
\ee
where ${\bf e}_i$ is the basis of the Cartesian coordinate system.
For $0<\alpha<1$ the operator 
${\partial^{\alpha,-}}/{\partial |x_i|^{\alpha}}$ 
is called the conjugate Riesz derivative \cite{Uch}.
Therefore, the operator
${\partial^{\alpha,-}}/{\partial |x_i|^{\alpha}}$
for all $\alpha>0$ can be called
the generalized conjugate Riesz derivative.

The Fourier transform ${\cal F}$ of 
the fractional derivative (\ref{CRFD}) is given by 
\be 
{\cal F} \left( \frac{\partial^{\alpha,-} f({\bf r})}{
\partial |x_i|^{\alpha}} \right)({\bf k}) 
= i \, k_i \, |k_i|^{\alpha-1} ({\cal F} f)({\bf k})
= i \, \operatorname{sgn}(k_i) \, 
|k_i|^{\alpha} ({\cal F} f)({\bf k}) . 
\ee
Using (\ref{s=2j}) and (\ref{CRFD}), we get
\be \label{s=2j+1}
\frac{\partial^{2j+1,-} f({\bf r})}{\partial |x_i|^{2j+1}} = 
(-1)^j \, 
\frac{\partial^{2j+1} f({\bf r})}{\partial x_i^{2j+1} } .
\ee
The fractional derivatives 
${\partial^{\alpha,-}}/{\partial |x_i|^{\alpha}}$
for odd orders $\alpha$ are local operators.
Note that the generalized conjugate Riesz derivative
${\partial^{2,-}}/{\partial |x_i|^{2}}$
cannot be considered as a derivative of second order
with respect to $|x_i|$.
The derivatives 
${\partial^{\alpha,-}}/{\partial |x_i|^{\alpha}}$
for even orders $\alpha=2j$ are non-local operators 
that cannot be considered
as usual derivatives ${\partial^{2j}}/{\partial x^{2j}_i}$.
For $\alpha=2$ the generalized conjugate Riesz derivative
is not the Laplacian.

Equations (\ref{s=2j}) and (\ref{s=2j+1}) allow us
to state that the usual local partial derivatives 
of integer orders are obtained from the operators 
${\partial^{\alpha,\pm}}/{\partial |x_i|^{\alpha}}$
in the following two cases:
(1) for odd values $\alpha=2j+1>0$
by ${\partial^{\alpha,-}}/{\partial |x_i|^{\alpha}}$ only; 
(2) for even values $\alpha=2j>0$ by 
${\partial^{\alpha,+}}/{\partial |x_i|^{\alpha}}$ only. 
The operators ${\partial^{\alpha,+}}/{\partial |x_i|^{\alpha}}$
with integer odd $\alpha=2j+1$ and
${\partial^{\alpha,-}}/{\partial |x_i|^{\alpha}}$
with integer even $\alpha=2j$, where $n \in \mathbb{N}$, 
are non-local operators.
Therefore we consider the lattice equations
with the lattice operators
$\mathbb{K}^{-} \left[ \alpha_i \atop i \right]$ and
$\mathbb{K}^{-,-} \left[ \alpha_i \, \beta_j \atop i \ j \right]$
as main lattice models to have the usual equations
with local spatial derivatives 
in the case $\alpha_i=\beta_i=1$ for all $i=1,2,3$.


\section{Fractional Fokker-Planck equation for continuum}

Using the statements (\ref{Limit-1}), 
(\ref{Limit-2}) and (\ref{Limit-3}), 
where $K^{-}_{\alpha}(n-m)$ are defined by (\ref{Kn1-}),
the lattice Fokker-Planck equation (\ref{LattEq-2}) 
are transformed by the combination 
${\cal F}^{-1} \circ \, \operatorname{Lim} \circ 
\, {\cal F}_{\Delta}$ 
into the fractional Fokker-Planck equation with
derivatives of non-integer orders 
with respect to space coordinates. 
This space-fractional Fokker-Planck equation
for the probability density $f({\bf r},t)$ has the form
\be \label{FFPE-1}
\frac{\partial f({\bf r},t)}{\partial t} = 
-\sum_{i=1}^3 D_i (\alpha) \, 
\frac{\partial^{\alpha_i,-}}{\partial |x_i|^{\alpha_i}} 
f({\bf r},t) + 
\frac{1}{2} \sum_{i=1}^3 \sum_{j=1}^3 D_{ij} (\alpha,\beta) \,
\frac{\partial^{\alpha_i,-}}{\partial |x_i|^{\alpha_i}}
\frac{\partial^{\beta_j,-}}{\partial |x_j|^{\beta_j}} \,
f ({\bf r},t) ,
\ee
where $D_i(\alpha)$ is the drift vector and 
$D_{ij}(\alpha,\beta)$ is the diffusion tensor
for the continuum that 
are defined by the lattice coupling constants
$g_{i}$ and $g_{ij}$ by the relations
\be \label{DiDij}
D_{i}(\alpha) = a^{\alpha_i}_i \, g_{i} , \quad
D_{ij} (\alpha,\beta) = 
2 \, a^{\alpha_i}_i \, a^{\beta_j} \, g_{ij} .
\ee

Using the definition (\ref{CRFD}),
the fractional Fokker-Planck equation (\ref{FFPE-1})
can be represented as the well-known continuity equation
\be \label{FFPE-2int}
\frac{\partial f({\bf r},t)}{\partial t} = 
-\sum_{i=1}^3 
\frac{\partial J_i ({\bf r},t)}{\partial x_i} ,
\ee
where $J_i$ is the probability flow
\be 
J_i ({\bf r},t) = 
\left\{
\aligned
D_i (\alpha) \, {\bf I}^{1-\alpha_i}_i f({\bf r},t) 
-\frac{1}{2} \sum_{j=1}^3 D_{ij} (\alpha,\beta) \,
{\bf I}^{1-\alpha_i}_i 
\frac{\partial^{\beta_j,-}}{\partial |x_j|^{\beta_j}}
f({\bf r},t) & 
\qquad 0<\alpha_i<1 , \\
D_i (\alpha) \, f({\bf r},t) 
-\frac{1}{2} \sum_{j=1}^3 D_{ij} (\alpha,\beta) \,
\frac{\partial^{\beta_j,-}}{\partial |x_j|^{\beta_j}}
f({\bf r},t) & 
\qquad \alpha_i=1 , \\
D_i (\alpha) \, \frac{\partial^{\alpha_i-1,+} f({\bf r},t)}{\partial |x_j|^{\alpha_i-1}} 
-\frac{1}{2} \sum_{j=1}^3 D_{ij} (\alpha,\beta) \,
\frac{\partial^{\alpha_i-1,+}}{\partial |x_j|^{\alpha_i-1}}
\frac{\partial^{\beta_j,-}}{\partial |x_j|^{\beta_j}}
f({\bf r},t) &
\qquad \alpha_i>1 , 
\endaligned
\right.
\ee
Note that coincidence of orders of fractional derivatives 
in the first and second terms allows us
to represent the fractional Fokker-Planck equation 
(\ref{FFPE-1}) in the form of 
the space-fractional continuity equation.
The fractional Fokker-Planck equation (\ref{FFPE-1})
can be represented as the fractional continuity equation
\be \label{FFPE-2}
\frac{\partial f({\bf r},t)}{\partial t} = 
-\sum_{i=1}^3 
\frac{\partial^{\alpha_i,-} J^{(frac)}_i ({\bf r},t)}{
\partial |x_i|^{\alpha_i}} ,
\ee
where $J^{(frac)}_i$ is the probability flow
\be \label{PF-1}
J^{(frac)}_i ({\bf r},t) = D_i (\alpha) \, f({\bf r},t) -
\frac{1}{2} \sum_{j=1}^3 D_{ij} (\alpha,\beta) \,
\frac{\partial^{\beta_j,-}}{\partial |x_j|^{\beta_j}}
f({\bf r},t) .
\ee
If $\alpha_i=1$, the continuity equation (\ref{FFPE-2})
has the standard form.


For one-dimensional case with $D_i (\alpha)=0$ and 
$f({\bf r},t)=f(x,t)$, 
equation (\ref{FFPE-1}) can be represented in the form
\be \label{FFPE-1-1d}
\frac{\partial f(x,t)}{\partial t} = 
K(\mu) \, \nabla^{\mu} f (x,t) ,
\ee
where $K(\mu)$ is the generalized diffusion constant,
\be 
K(\mu) = \frac{1}{2} D_{11} (\alpha,\beta) , 
\ee
and $\nabla^{\mu}$ is the fractional derivative of order $\mu$, 
\be \label{nabla-1}
\nabla^{\mu} =
\frac{\partial^{\alpha_1,-}}{\partial |x|^{\alpha_1}}
\frac{\partial^{\beta_1,-}}{\partial |x|^{\beta_1}} , 
\quad 
\mu =\alpha_1+\beta_1 .
\ee
Note that for sufficiently good functions, 
the operator (\ref{nabla-1}) can be represented in the form
$\nabla^{\mu} = {\partial^{\mu,+}}/{\partial |x|^{\mu}}$,
but it cannot be done in the general case.
Equation (\ref{FFPE-1-1d}) describes the fractional diffusion
processes with the Poissonian waiting time and 
the L\'evy distribution for the jump length
(see Section 3.5 of \cite{MK2000}).
In \cite{MK2000} the space-fractional diffusion 
equation (\ref{FFPE-1-1d}) contains 
the Weyl fractional derivative $\nabla^{\mu}$ of order $\mu$, 
which is equivalent to the Riesz operator 
${\partial^{\mu,+}}/{\partial |x|^{\mu}}$ 
in one dimension.
The solution of equation (\ref{FFPE-1-1d}) can be obtained 
analytically by using the Fox function $H^{1,1}_{2,2}$
(for details see Section 3.5 in \cite{MK2000} 
and \cite{MainardiPS}).
The exact calculation of fractional moments \cite{MK2000} gives
\be \label{x-delta-2}
\langle |x(t)|^{\delta} \rangle = 
\frac{2 \, (K(\mu))^{\delta / \mu} \, \Gamma (- \delta /\mu) \, \Gamma (1+\delta)}{
\mu \, \Gamma (-\delta/2) \, \Gamma (1+\delta/2)} \, t^{\delta / \mu} ,
\ee
where $0< \delta <\mu \le 2$.


The time-fractional lattice Fokker-Planck equation 
(\ref{LattEq-2m}) are transformed by the combination 
${\cal F}^{-1} \circ \, \operatorname{Lim} \circ 
\, {\cal F}_{\Delta}$ 
into the space-time fractional Fokker-Planck equation 
\be \label{FFPE-1m}
\frac{\partial f({\bf r},t)}{\partial t} = 
-\sum_{i=1}^3 D_i (\alpha) \, 
\frac{\partial^{\alpha_i,-}}{\partial |x_i|^{\alpha_i}} 
f({\bf r},t) + 
\frac{1}{2} \sum_{i=1}^3 \sum_{j=1}^3 
D_{ij} (\alpha,\beta,\gamma) \,
\frac{\partial^{\alpha_i,-}}{\partial |x_i|^{\alpha_i}}
\frac{\partial^{\beta_j,-}}{\partial |x_j|^{\beta_j}} \,
_0^{RL}D^{1-\gamma}_t \, f ({\bf r},t) ,
\ee
where $_0^{RL}D^{1-\gamma}_t$ is the Riemann-Liouville 
fractional derivative with respect to time
that describes the power-law memory.
For one-dimensional case with $D_i (\alpha)=0$ and 
$f({\bf r},t)=f(x,t)$, 
equation (\ref{FFPE-1m}) can be represented in the form
\be \label{FFPE-1m-1d}
\frac{\partial f(x,t)}{\partial t} = 
\, _0^{RL}D^{1-\gamma}_t \, 
K(\mu , \gamma) \, \nabla^{\mu} f (x,t) ,
\ee
where
\be
K(\mu , \gamma) = \frac{1}{2} D_{11} (\alpha,\beta,\gamma) , 
\ee
and the fractional derivative $\nabla^{\mu}$
is defined by (\ref{nabla-1}).
Equation (\ref{FFPE-1m-1d}) describes a random walk characterized 
by waiting time and jump length 
(see Section 3.6 in \cite{MK2000}).
The competition between long rests (waiting events) 
and long jumps (motion events) in the L\'evy walks processes 
is given \cite{ZK1995} as
\be 
\langle x^2(t) \rangle \sim
\left\{
\begin{array}{cc}
t^{2 +\gamma - \mu} 
& 0< \gamma <1, \\
& \\
t^{3 - \mu} 
& \gamma >1 ,
\end{array}
\right.
\ee
where $1<\mu<2$. It should be noted that 
the continuum form of the L\'evy flights
is described by the drift term with 
the first order derivative ($\alpha_i=1$)
as proposed in \cite{Fogedby} 
and derived from the continuous time random walk in \cite{MBK-1}.
The solutions of Cauchy problems for the space-time 
fractional diffusion equation with 
the Riesz-Feller fractional derivatives
are described in \cite{MLP2001}.


The time-fractional lattice Fokker-Planck equation 
(\ref{LattEq-2mm}) is transformed by the combination 
${\cal F}^{-1} \circ \, \operatorname{Lim} \circ 
\, {\cal F}_{\Delta}$ 
into the space-time fractional continuum 
Fokker-Planck equation 
\be \label{FFPE-1mm}
\frac{\partial f({\bf r},t)}{\partial t} = \,
_0^{RL}D^{1-\gamma}_t \, {\cal L}^{\alpha, \beta}_{CFP}
\, f ({\bf r},t) ,
\ee
where 
\be
{\cal L}^{\alpha, \beta}_{CFP} =
{\cal F}^{-1} \circ \, \operatorname{Lim} \circ 
\, {\cal F}_{\Delta} \left( 
{\cal L}^{\alpha, \beta}_{LFP}
\right) 
\ee
is the continuum Fokker-Planck operatorof the form
\be
{\cal L}^{\alpha, \beta}_{CFP} =
-\sum_{i=1}^3 D_i (\alpha) \, 
\frac{\partial^{\alpha_i,-}}{\partial |x_i|^{\alpha_i}} + 
\frac{1}{2} \sum_{i=1}^3 \sum_{j=1}^3 
D_{ij} (\alpha,\beta,\gamma) \,
\frac{\partial^{\alpha_i,-}}{\partial |x_i|^{\alpha_i}}
\frac{\partial^{\beta_j,-}}{\partial |x_j|^{\beta_j}} .
\ee
For $\alpha_i=\beta_i=1$, equation (\ref{FFPE-1mm})
takes the form of the time-fractional
Fokker-Planck equation that 
is suggested in \cite{MBK-1,MBK-2,BMK}.


If $\alpha_i=\beta_i=\gamma=1$ for all $i=1,2,3$, then 
equations (\ref{FFPE-1}), (\ref{FFPE-1m}) and (\ref{FFPE-1mm}) 
for the probability density $f({\bf r},t)$ give
the well-known Fokker-Planck equation in the form
\be
\frac{\partial f({\bf r},t)}{\partial t} = 
-\sum_{i=1}^3 D_i \, \frac{\partial}{\partial x_i} f({\bf r},t) + 
\frac{1}{2}\sum_{i=1}^3 \sum_{j=1}^3 D_{ij} \,
\frac{\partial^2}{\partial x_i \, \partial x_j} f({\bf r},t) ,
\ee
where $D_i=D_i(1)$ is the usual drift vector and 
$D_{ij}=D_{ij}(1,1)$ is the usual diffusion tensor.


\section{Conclusion}

In this paper three-dimensional lattice models with 
long-range drift and diffusion of particles are suggested.
These proposed lattice models can be considered
as a possible microscopic basis to describe 
the anomalous diffusion in continuum.
The suggested type of lattice long-range drift and diffusion 
can be considered for non-integer (fractional) values 
of the parameters $\alpha_i$, $\beta_i$, $\gamma$. 
This allows us to have lattice equations for 
the fractional nonlocal diffusion and transport processes.
The proposed forms of the drift and diffusion of particles 
in lattice allow us to obtain
the continuum equations with the generalized conjugate
Riesz derivatives of fractional orders 
by using the approaches and methods 
proposed in \cite{JMP2006,JPA2006}.
The suggested three-dimensional models with 
long-range lattice drift and diffusion of 
the types (\ref{Kn1+}) and (\ref{Kn1-}) 
can be considered as lattice analogs of 
the fractional diffusion and drift in nonlocal continuum. 
Different fractional generalizations of Fokker-Planck equation 
for continuum can be obtained by using the suggested 
lattice approach.
We expect that the proposed three-dimensional lattice 
Fokker-Planck equations
can play an important role in the description
of nonlocal processes in microscale and nanoscale 
because at these scales the interatomic interactions 
can be prevalent in determining the properties of media.

Let us note some possible generalizations of 
the proposed lattice models.
We assume that the suggested lattice Fokker-Planck equations 
can be generalized in the form of lattice Kramers-Moyal 
equation for the case of the high-order terms 
by using the different fractional-order derivatives.
The suggested lattice models can be generalized for
the lattices with dislocation and disclinations
that are connected with non-commutativity of
the lattice operators (\ref{Ds1}).
In this paper, we consider the primitive 
orthorhombic Bravais lattice for simplification.
It is interesting to generalize the suggested 
consideration for other type of Bravais lattices
such as triclinic, monoclinic, rhombohedral and hexagonal.
The suggested models of unbounded lattices
can be generalized for the bounded physical lattices.
We also assume that the proposed lattice approach
to the fractional diffusion 
can be generalized for lattices,
which are characterized by fractal spatial dispersion
\cite{JPA2008,FL-2,FL-3},
and correspondent models 
for fractal media \cite{SP1985,MGN1994} 
(see also \cite{CNSNS2014,JMP2014,CSF2014}).

We also can note some remaining challenges and questions
in the suggested approach to fractional diffusion. 
The function $f({\bf n},t)$ has the meaning of 
probability density on the lattice, and 
it should be positively defined.
It is known that this condition for continuum case
leads to restriction for the parameters 
$\alpha$, $\beta$, $\gamma$, $\mu$.
For example, we have the condition $0< \mu \le 2$ 
for L\'evy processes on continuum.
A rigorous consideration of positiveness of $f({\bf n},t)$
for set of these parameter does not exist at the present time.
Exact mathematical conditions of existence
of solutions for the lattice Fokker-Planck equation 
can be important to describe anomalous long-range particle 
drift and diffusion on three-dimensional physical lattices.



\section*{Appendix: Riesz fractional integral}

The Riesz fractional integration is defined by
\be \label{RInt}
{\bf I}^{\alpha}_x f(x) = 
{\cal F}^{-1} \Bigl( |k|^{-\alpha} ({\cal F} f)(k) \Bigr) .
\ee
The fractional integration (\ref{RInt}) can be realized in the form of 
the Riesz potential defined as the Fourier's convolution of the form
\be \label{RFP}
{\bf I}^{\alpha}_x f(x)= 
\int_{\mathbb{R}^n} R_{\alpha}(x-z) f(z) dz, \quad (\alpha >0) ,
\ee
where the function $R_{\alpha}(x)$ is the Riesz kernel.
If $\alpha>0$, the function $R_{\alpha}(x)$ is defined by 
\be 
R_{\alpha}(x) = 
\left\{
\begin{array}{cc}
\gamma^{-1}_n(\alpha) |x|^{\alpha-n} 
& \alpha \ne n+2k, \\
& \\
- \gamma^{-1}_n(\alpha) |x|^{\alpha-n} \ln |x| 
& \alpha =n+2k ,
\end{array}
\right.
\ee
where $n \in \mathbb{N}$, and
the constant $\gamma_n(\alpha)$ has the form
\be 
\gamma_n(\alpha)=
\left\{
\begin{array}{cc}
2^{\alpha} \pi^{n/2}\Gamma(\alpha/2)/ \Gamma(\frac{n-\alpha}{2}) &
\alpha \ne n+2k, \\
& \\
(-1)^{(n-\alpha)/2}2^{\alpha-1} \pi^{n/2} \;
\Gamma(\alpha/2) \; 
\Gamma( 1+[\alpha-n]/2) 
& \alpha =n+2k .
\end{array}
\right.
\ee
The Fourier transform of the Riesz fractional integration is given by
\be \label{FTRI}
{\cal F} \Bigl( {\bf I}^{\alpha}_x f(x)\Bigr) = 
|k|^{-\alpha} ({\cal F} f)(k) . 
\ee




\begin{thebibliography}{**}



\vspace{-3mm} \bibitem{Risken} H. Risken, 
{\it The Fokker-Planck Equation}
(Springer, Berlin, 1984).


\vspace{-3mm} \bibitem{BG1990} J.P. Bouchaud, A. Georges,
"Anomalous diffusion in disordered media: 
statistical mechanisms, models and physical applications",
Physics Reports. Vol.195. No.4-5. (1990) 127-293.

\vspace{-3mm} \bibitem{SZK1993} M.F. Shlesinger, G.M. Zaslavsky, J. Klafter,
"Strange kinetics", Nature. Vol.363. No.6424. (1993) 31-37.

\vspace{-3mm} \bibitem{MK2000} R. Metzler, J. Klafter, 
"The random walk's guide to anomalous diffusion: 
a fractional dynamics approach", 
Physics Reports. Vol.339. No.1. (2000) 1-77.


\vspace{-3mm} \bibitem{Hughes} B.D. Hughes, 
{\it Random Walks and Random Environments. 
Vol. 1: Random Walks} (Oxford Univ. Press, 1995); 
Vol. 2: Random Environments. (Oxford Univ. Press, 1996).


\vspace{-3mm} \bibitem{SKM} S.G. Samko, A.A. Kilbas, O.I. Marichev,
{\it Integrals and Derivatives of Fractional Order and Applications} 
(Nauka i Tehnika, Minsk, 1987); and
{\it Fractional Integrals and Derivatives Theory and Applications} 
(Gordon and Breach, New York, 1993). 

\vspace{-3mm} \bibitem{KST} A.A. Kilbas, H.M. Srivastava, J.J. Trujillo,
{\it Theory and Applications of Fractional 
Differential Equations} (Elsevier, Amsterdam, 2006). 

\vspace{-3mm} \bibitem{Ort} M.D. Ortigueira,
{\it Fractional Calculus for Scientists and Engineers} 
(Springer, Netherlands, 2011).

\vspace{-3mm} \bibitem{Uch} V.V. Uchaikin, 
{\it Fractional Derivatives for Physicists and Engineers. 
Volume I. Background and Theory} 
(Springer, Higher Education Press, 2012).

\vspace{-3mm} \bibitem{Zhou} Y. Zhou,
{\it Basic Theory of Fractional Differential Equations} 
(World Scientific, Singapore, 2014).

\vspace{-3mm} \bibitem{Mainardi1997} F. Mainardi,
"Fractional calculus: Some basic problems 
in continuum and statistical mechanics",
in the book A. Carpinteri, F. Mainardi (Eds.), 
{\it Fractals and Fractional Calculus in Continuum Mechanics} 
(Springer, Wien and New York, 1997) 291-348. (arXiv:1201.0863)

\vspace{-3mm} \bibitem{FC1} R.E. Gutierrez, J.M. Rosario, J.A. Tenreiro Machado,
"Fractional order calculus: 
Basic concepts and engineering applications",
Mathematical Problems in Engineering. 
Vol.2010. (2010) 375858. 

\vspace{-3mm} \bibitem{FC2} D. Valerio, J.J. Trujillo, M. Rivero, 
J.A. Tenreiro Machado, D. Baleanu,
"Fractional calculus: A survey of useful formulas",
The European Physical Journal. Special Topics.
Vol.222. No.8. (2013) 1827-1846.


\vspace{-3mm} \bibitem{MLP2001} 
F. Mainardi, Yu. Luchko, G. Pagnini,
"The fundamental solution of 
the space-time fractional diffusion equation",
Fractional Calculus and Applied Analysis. 
Vol.4. No.2. (2001) 153-192. (arXiv:cond-mat/0702419)

\vspace{-3mm} \bibitem{Zaslavsky2002} G.M. Zaslavsky, 
"Chaos, fractional kinetics, and anomalous transport", 
Physics Reports. Vol.371. No.6. (2002) 461-580.

\vspace{-3mm} \bibitem{MK2004} R. Metzler, J. Klafter, 
"The restaurant at the end of the random walk: recent developments 
in the description of anomalous transport by fractional dynamics", 
Journal of Physics A. Vol.37. No.31. (2004) R161-R208.

\vspace{-3mm} \bibitem{KLM} J. Klafter, S.C. Lim, R. Metzler (Eds.), 
{\it Fractional Dynamics. Recent Advances} 
(World Scientific, Singapore, 2011).

\vspace{-3mm} \bibitem{MS2012} M.M. Meerschaert, A. Sikorskii,
{\it Stochastic Models for Fractional Calculus} 
(Walter de Gruyter, Berlin/Boston, 2012).

\vspace{-3mm} \bibitem{US} V. Uchaikin, R. Sibatov,
{\it Fractional Kinetics in Solids: Anomalous Charge 
Transport in Semiconductors, Dielectrics and Nanosystems} 
(World Scientific, Singapore, 2013).


\vspace{-3mm} \bibitem{CM} A. Carpinteri, F. Mainardi (Eds.),
{\it Fractals and Fractional Calculus in Continuum Mechanics} 
(Springer, New York, 1997).

\vspace{-3mm} \bibitem{Hilfer} R. Hilfer (Ed.), 
{\it Applications of Fractional Calculus in Physics} 
(World Scientific, Singapore, 2000).

\vspace{-3mm} \bibitem{SATM} J. Sabatier, O.P. Agrawal, J.A. Tenreiro Machado (Eds.)
{\it Advances in Fractional Calculus. 
Theoretical Developments and Applications in Physics and Engineering}
(Springer, Dordrecht, 2007).

\vspace{-3mm} \bibitem{LA} A.C.J. Luo, V.S. Afraimovich (Eds.),
{\it Long-range Interaction, Stochasticity and 
Fractional Dynamics} (Springer, Berlin, 2010).

\vspace{-3mm} \bibitem{Mainardi} F. Mainardi, 
{\it Fractional Calculus and Waves in Linear Viscoelasticity: 
An Introduction to Mathematical Models} 
(World Scientific, Singapore, 2010).

\vspace{-3mm} \bibitem{TarasovSpringer} V.E. Tarasov, 
{\it Fractional Dynamics: Applications of Fractional Calculus 
to Dynamics of Particles, Fields and Media} 
(Springer, New York, 2011).

\vspace{-3mm} \bibitem{IJMPB2013} V.E. Tarasov,
"Review of some promising fractional physical models",
International Journal of Modern Physics B. 
Vol.27. No.9. (2013) 1330005. (arXiv:1502.07681)

\vspace{-3mm} \bibitem{APSS2014a}
T.M. Atanackovic, S. Pilipovic, B. Stankovic, D. Zorica, 
{\it Fractional Calculus with Applications in Mechanics: 
Vibrations and Diffusion Processes} 
(Wiley-ISTE, London, Hoboken, 2014).

\vspace{-3mm} \bibitem{APSS2014b}
T.M. Atanackovic, S. Pilipovic, B. Stankovic, D. Zorica, 
{\it Fractional Calculus with Applications in Mechanics: 
Wave Propagation, Impact and Variational Principles} 
(Wiley-ISTE, London, Hoboken, 2014).


\vspace{-3mm} \bibitem{Zaslavsky1994} G.M. Zaslavsky, 
"Fractional kinetic equation for Hamiltonian chaos", 
Physica D. Vol.76. No.1-3. (1994) 110-122.

\vspace{-3mm} \bibitem{SZ1997} A.I. Saichev, G.M. Zaslavsky, 
"Fractional kinetic equations: solutions and applications", 
Chaos. Vol.7. No.4. (1997) 753-764.

\vspace{-3mm} \bibitem{Milovanov2001} A.V. Milovanov, 
"Stochastic dynamics from the fractional 
Fokker-Planck-Kolmogorov equation: 
Large-scale behavior of the turbulent transport coefficient", 
Physical Review E. Vol.63. No.4. (2001) 047301.

\vspace{-3mm} \bibitem{MN2002} R. Metzler, T.F. Nonnenmacher, 
"Space- and time-fractional diffusion and wave equations, 
fractional Fokker-Planck equations, and physical motivation",
Chemical Physics. Vol.284. No.1-2. (2002) 67-90.


\vspace{-3mm} \bibitem{Chaos2005} V.E. Tarasov,
"Fractional Fokker-Planck equation for fractal media",
Chaos. Vol.15. No.2. (2005) 023102. (arXiv:nlin.CD/0602029) 

\vspace{-3mm} \bibitem{Physica2008} V.E. Tarasov, G.M. Zaslavsky,
"Fokker-Planck equation with fractional coordinate derivatives",
Physica A. Vol.387. No.26. (2008) 6505-6512. (arXiv:0805.0606)

\vspace{-3mm} \bibitem{TSMD2012} Z. Tomovski, T. Sandev, R. Metzler, J. Dubbeldam,
"Generalized space-time fractional diffusion equation with composite fractional time derivative",
Physica A. Vol.391. No.8. (2012) 2527-2542.



\vspace{-3mm} \bibitem{MBK-1} R. Metzler, E. Barkai, J. Klafter, 
from a generalised master equation,
"Deriving fractional Fokker-Planck equations 
from a generalised master equation",
Europhysics Letters. Vol.46. No.4. (1999) 431-436.

\vspace{-3mm} \bibitem{MBK-2} R. Metzler, E. Barkai, J. Klafter, 
"Anomalous diffusion and relaxation close to thermal equilibrium: 
A fractional Fokker-Planck equation approach",
Physical Review Letters. Vol.82. No.18. (1999) 3563-3567.

\vspace{-3mm} \bibitem{BMK} E. Barkai, R. Metzler, J. Klafter, 
"From continuous time random walks to 
the fractional Fokker-Planck equation",
Physical Review E. Vol.61. No.1. (2000) 132-138.

\vspace{-3mm} \bibitem{Fogedby} H.C. Fogedby, 
"L\'evy flights in random environments",
Physical Review Letters. 
Vol.73. No.9. (1994) 2517-2520.


\vspace{-3mm} \bibitem{LFPE-1} S. Succi, S. Melchionna, J.-P. Hansen, 
"Lattice Fokker-Planck equation",
International Journal of Modern Physics C.
Vol.17. No.4. (2006) 459-470.

\vspace{-3mm} \bibitem{LFPE-2} 
D. Moroni, B. Rotenberg, J.-P. Hansen, S. Succi, S. Melchionna,
"Solving the Fokker-Planck kinetic equation on a lattice",
Physical Review E. Vol.73. No.6. (2006) 066707.
(arXiv:cond-mat/0512497)

\vspace{-3mm} \bibitem{LFPE-3} F. Wu, W. Shi, F. Liu,
"A lattice Boltzmann model for the Fokker-Planck equation",
Communications in Nonlinear Science and Numerical Simulation. 
Vol.17. No.7. (2012) 2776-2790.

\vspace{-3mm} \bibitem{LFPE-4} D. Moroni, J.-P. Hansen, S. Melchionna, S. Succi,
"On the use of lattice Fokker-Planck models for hydrodynamics",
Europhysics Letters. Vol.75. No.3. (2006) 399-405.

\vspace{-3mm} \bibitem{LFPE-5} S. Melchionna, S. Succi, J.-P. Hansen, 
"Simulation of single-file ion transport with 
the lattice Fokker-Planck equation",
Physical Review E. Vol.73. No.1. (2006) 017701.

\vspace{-3mm} \bibitem{LFPE-6} S. Singh, G. Subramanian, S. Ansumali,
"Lattice Fokker Planck for dilute polymer dynamics",
Physical Review E. Vol.88. No.1. (2013) 013301.


\vspace{-3mm} \bibitem{Sedov} L.I. Sedov, {\it A course in Continuum Mechanics} 
Vol.1-4. (Wolters-Noordhoff Publishing, Netherlands, 1971).

\vspace{-3mm} \bibitem{Born} M. Born, K. Huang, 
{\it Dynamical Theory of Crystal Lattices} , 
(Oxford University Press, Oxford, 1954). 

\vspace{-3mm} \bibitem{MMW} A.A. Maradudin, E.W. Montroll, G.H. Weiss,
{\it Theory of Lattice Dynamics in the Harmonic Approximation} 
(Academic Press, New York, 1963). 

\vspace{-3mm} \bibitem{Bo} H. B\"otteger,
{\it Principles of the Theory of Lattice Dynamics} 
(Academie-Verlag, Berlin, 1983). 

\vspace{-3mm} \bibitem{Kosevich} A.M. Kosevich, 
{\it The Crystal Lattice. Phonons, Solitons, 
Dislocations, Superlattices} , 
Second Edition (Wiley-VCH, Berlin, New York, 2005). 


\vspace{-3mm} \bibitem{Bogolyubov1} N.N. Bogolyubov, N.M. Krylov, 
"On the Fokker - Planck equation, 
which appear in the perturbation method based 
on the spectral properties of the perturbed Hamiltonian", 
Notes Department of Mathematical Physics. 
Institute of Nonlinear Mechanics. 
Academy of Sciences of the Ukrainian SSR. 
Vol.4. (1939) 5-80. in Ukrainian.

\vspace{-3mm} \bibitem{Bogolyubov2} N.N. Bogolyubov, 
{\it Collected Works in 12 volumes} . 
Volume 5: {\it Non-equilibrium statistical mechanics, 1939-1980}.
(Nauka, Moscow, 2006). in Russian.


\vspace{-3mm} \bibitem{LLI-1} A. Campa, T. Dauxois, S. Ruffo,
"Statistical mechanics and dynamics of solvable 
models with long-range interactions",
Physics Reports. Vol.480. No.3-6. (2009) 57-159. 
(arXiv:0907.0323)

\vspace{-3mm} \bibitem{LLI-2} J. Barre, F. Bouchet, T. Dauxois, S. Ruffo,
"Large deviation techniques applied 
to systems with long-range interactions", 
Journal of Statistical Physics. 
Vol.119. No.3-4. (2005) 677-713.
(arXiv:cond-mat/0406358)

\vspace{-3mm} \bibitem{LLI-3} S. Ruffo,
"Equilibrium and nonequilibrium properties of 
systems with long-range interactions",
The European Physical Journal B. 
Vol.64. No.3-4. (2008) 355-363. (arXiv:0711.1173)

\vspace{-3mm} \bibitem{LLI-4} 
C. Nardini, S. Gupta, S. Ruffo, T. Dauxois, F. Bouchet,
"Kinetic theory for non-equilibrium stationary states 
in long-range interacting systems",
Journal of Statistical Mechanics. 
Vol.2012. No.1. (2012) L01002. (arXiv:1111.6833)

\vspace{-3mm} \bibitem{LLI-5} 
Y. Levin, R. Pakter, F.B. Rizzato, T,N. Teles, F.P.C. Benetti,
"Nonequilibrium statistical mechanics of systems 
with long-range interactions",
Physics Reports. Vol.535. No.1. (2014) 1-60.
(arXiv:1310.1078)

\vspace{-3mm} \bibitem{LLI-6} H. Hinrichsen, 
"Non-equilibrium phase transitions with long-range interactions",
Journal of Statistical Mechanics. Vol.2007. No.7. (2007) P07006.
(arXiv:cond-mat/0702169)

\vspace{-3mm} \bibitem{LLI-7}
R. Bachelard, C. Chandre, D. Fanelli, X. Leoncini, S. Ruffo,
"Abundance of regular orbits and nonequilibrium phase transitions 
in the thermodynamic limit for long-range systems",
Physics Review Letters. Vol.101. No.26. (2008) 260603. 


\vspace{-3mm} \bibitem{JMP2006} V.E. Tarasov,
"Map of discrete system into continuous",
Journal of Mathematical Physics. 
Vol.47. No.9. (2006) 092901. (arXiv:0711.2612) 

\vspace{-3mm} \bibitem{JPA2006} V.E. Tarasov,
"Continuous limit of discrete systems with long-range interaction",
Journal of Physics A. Vol.39. No.48. (2006) 14895-14910. (arXiv:0711.0826) 


\vspace{-3mm} \bibitem{Chaos2006} V.E. Tarasov, G.M. Zaslavsky,
"Fractional dynamics of coupled oscillators 
with long-range interaction",
Chaos. Vol.16. No.2. (2006) 023110. (arXiv:nlin.PS/0512013) 

\vspace{-3mm} \bibitem{CNSNS2006} V.E. Tarasov, G.M. Zaslavsky,
"Fractional dynamics of systems with long-range interaction",
Communications in Nonlinear Science and Numerical Simulation. 
Vol.11. No.8. (2006) 885-898. (arXiv:1107.5436) 

\vspace{-3mm} \bibitem{Laskin} N. Laskin, G.M. Zaslavsky, 
"Nonlinear fractional dynamics on a lattice 
with long-range interactions",
Physica A. Vol.368. No.1. (2006) 38-54. (arXiv:nlin.SI/0512010)


\vspace{-3mm} \bibitem{CEJP2013}  V.E. Tarasov,
"Lattice model with power-law spatial dispersion for fractional elasticity",
Central European Journal of Physics. 
Vol.11. No.11. (2013) 1580-1588.
(arXiv:1501.01201)

\vspace{-3mm} \bibitem{MOM2014} V.E. Tarasov,
"Lattice model of fractional gradient and integral elasticity: 
Long-range interaction of Grunwald-Letnikov-Riesz type",
Mechanics of Materials. Vol.70. No.1. (2014) 106-114.
(arXiv:1502.06268)

\vspace{-3mm} \bibitem{MPLB2014} V.E. Tarasov, 
"General lattice model of gradient elasticity",
Modern Physics Letters B. 
Vol.28. No.7. (2014) 1450054. (arXiv:1501.01435)

\vspace{-3mm} \bibitem{ISRN-CMP2014} VV.E. Tarasov, 
"Fractional gradient elasticity from spatial dispersion law",
ISRN Condensed Matter Physics. 
Vol.2014. (2014) 794097. (arXiv:1306.2572) 

\vspace{-3mm} \bibitem{IJSS2014}.E. Tarasov, 
"Lattice with long-range interaction of power-law type 
for fractional non-local elasticity",
International Journal of Solids and Structures.
Vol.51. No.15-16. (2014) 2900-2907. (arXiv:1502.05492)

\vspace{-3mm} \bibitem{ND2014} V.E. Tarasov,
"Non-linear fractional field equations: 
weak non-linearity at power-law non-locality",
Nonlinear Dynamics. (2015) accepted for publication.
DOI: 10.1007/s11071-014-1342-0


\vspace{-3mm} \bibitem{Prudnikov} A.P. Prudnikov, Yu.A. Brychkov, O.I. Marichev,
{\it Integrals and Series. Vol. 1: Elementary Functions} 
(Gordon and Breach, New York, 1986).

\vspace{-3mm} \bibitem{Erdelyi} A. Erdelyi, W. Magnus, 
F. Oberhettinger, F.G. Tricomi,
{\it Higher Transcendental Functions} Vol.1.
(McGraw-Hill, New York, 1953), and 
(Krieeger, Melbourne, Florida, 1981).


\vspace{-3mm} \bibitem{MainardiPS} F. Mainardi, 
G. Pagnini, R.K. Saxena,
"Fox H functions in fractional diffusion",
Journal of Computational and Applied Mathematics.
Vol.178. No.1-2. (2005) 321-331.


\vspace{-3mm} \bibitem{ZK1995} G. Zumofen, J. Klafter, 
"Laminar–localized-phase coexistence in dynamical systems",
Physical Review E. Vol.51. No.3. (1995) 1818-1821.


\vspace{-3mm} \bibitem{JPA2008} V.E. Tarasov,
"Chains with fractal dispersion law",
Journal of Physics A. Vol.41. No.3. (2008) 035101. (arXiv:0804.0607)
 
\vspace{-3mm} \bibitem{FL-2} T.M. Michelitsch, G.A. Maugin, F.C.G.A. Nicolleau, A.F. Nowakowski, S. Derogar,
"Dispersion relations and wave operators in 
self-similar quasicontinuous linear chains",
Physical Review E. Vol.80. No.1. (2009) 011135. (arXiv:0904.0780)

\vspace{-3mm} \bibitem{FL-3} T.M. Michelitsch, G.A. Maugin, 
F.C.G.A. Nicolleau, A.F. Nowakowski, S. Derogar,
"Wave propagation in quasi-continuous linear chains with 
self-similar harmonic interactions: 
Towards a fractal mechanics", in
{\it Mechanics of Generalized Continua:
Advanced Structured Materials} 
Vol.7. (Springer, Berlin, 2011) Chapter 11. pp.231-244.



\vspace{-3mm} \bibitem{SP1985} B.O. Shaughnessy, I. Procaccia, 
"Analytical solutions for diffusion on fractal objects", 
Physical Review Letters. Vol.54. No.5. (1985) 455-458.

\vspace{-3mm} \bibitem{MGN1994} R. Metzler, W.G. G1ockle, T.F. Nonnenmacher,,
"Fractional model equation for anomalous diffusion",
Physica A. Vol.211. No.1. (1994) 13-24.

\vspace{-3mm} \bibitem{CNSNS2014} V.E. Tarasov,
"Vector calculus in non-integer dimensional space and 
its applications to fractal media",
Communications in Nonlinear Science and Numerical Simulation.
Vol.20. No.2. (2015) 360-374. (arXiv:1503.02022)

\vspace{-3mm} \bibitem{JMP2014} V.E. Tarasov,
"Anisotropic fractal media by vector calculus 
in non-integer dimensional space",
Journal of Mathematical Physics. 
Vol.55. No.8. (2014) 083510. (arXiv:1503.02392)

\vspace{-3mm} \bibitem{CSF2014} V.E. Tarasov,
"Flow of Fractal Fluid in Pipes:
Non-Integer Dimensional Space Approach", 
Ñhaos, Solitons and Fractals. Vol.67. (2014) 26-37.
(arXiv:1503.02842)


\end{thebibliography}
\end{document}